\documentclass[prb,twocolumn,showpacs,floatfix,amsmath,amssymb,superscriptaddress]{revtex4}
\usepackage[pdftex]{graphicx}
\usepackage{latexsym}
\usepackage{graphicx}
\usepackage{times}
\usepackage{amsmath}
\usepackage{dcolumn}
\usepackage{latexsym,amsmath,amssymb,bm,euscript}
\begin{document}

\title{Non-Abelian statistics and topological quantum information processing in 1D wire networks}

\author{Jason Alicea}
\affiliation{Department of Physics, California Institute of Technology,
Pasadena, California 91125}
\affiliation{Department of Physics and Astronomy, University of California, Irvine, California 92697}

\author{Yuval Oreg}
\affiliation{Department of Condensed Matter Physics, Weizmann Institute of Science, Rehovot, 76100, Israel}

\author{Gil Refael}
\affiliation{Department of Physics, California Institute of Technology,
Pasadena, California 91125}

\author{Felix von Oppen}
\affiliation{Dahlem Center for Complex Quantum Systems and Fachbereich Physik,
Freie Universit\"at Berlin, 14195 Berlin, Germany}

\author{Matthew P. A. Fisher}
\affiliation{Department of Physics, California Institute of Technology,
Pasadena, California 91125}

\date{\today}

\begin{abstract}
  Topological quantum computation provides an elegant way around decoherence, as one encodes quantum information in a non-local fashion that the environment finds difficult to corrupt.  Here we establish that one of the key operations---braiding of non-Abelian anyons---can be implemented in \emph{one-dimensional} semiconductor wire networks.  Previous work\cite{1DwiresLutchyn,1DwiresOreg} provided a recipe for driving semiconducting wires into a topological phase supporting long-sought particles known as Majorana fermions that can store topologically protected quantum information.  Majorana fermions in this setting can be transported, created, and fused by applying locally tunable gates to the wire.  More importantly, we show that networks of such wires allow braiding of Majorana fermions and that they exhibit non-Abelian statistics like vortices in a $p+ip$ superconductor.  We propose experimental setups that enable the Majorana fusion rules to be probed, along with networks that allow for efficient exchange of arbitrary numbers of Majorana fermions.  This work paves a new path forward in topological quantum computation that benefits from physical transparency and experimental realism.
\end{abstract}

\maketitle

%%%%%%%%%%%%%%%%%%%%%%%%%%%%%%%%%%%%%%%%%%%%%%%%%%%%%%%%%%%%%%%%%%%%%%

The experimental realization of a quantum computer ranks among the foremost outstanding problems in condensed matter physics, particularly in light of the revolutionary rewards the achievement of this goal promises.  Typically, decoherence presents the primary obstacle to fabricating a scalable quantum computer.  In this regard \emph{topological} quantum computing holds considerable promise, as here one embeds quantum information in a non-local, intrinsically decoherence-free fashion\cite{kitaev,Freedman98,Freedman03,TopologicalQubits,MeasurementOnlyTQC,TQCreview}.  The core ideas can be illustrated with a toy model of a spinless, two-dimensional (2D) $p+ip$ superconductor.  Vortices in such a state bind exotic particles known as Majorana fermions, which cost no energy and therefore generate a ground state degeneracy.  Because of the Majoranas, vortices exhibit non-Abelian braiding statistics\cite{Leinaas,Fredenhagen,Froehlich,ReadGreen,Ivanov}: adiabatically exchanging vortices noncommutatively transforms the system from one ground state to another.  Quantum information encoded in this ground state space can be controllably manipulated by braiding operations---something the environment finds difficult to achieve.

Despite this scheme's elegance, realizing suitable topological phases poses a serious challenge.  Most effort has focused on the quantum Hall state at filling fraction\cite{MooreRead,ReadGreen} $\nu = 5/2$, though very recently the list of candidate experimental systems has rapidly expanded.  Indeed, topological insulators\cite{FuKane,Linder}, semiconductor heterostructures\cite{Sau,Alicea}, non-centrosymmetric superconductors\cite{SatoFujimoto,PatrickProposal,Ghosh}, and quantum Hall systems at integer plateau transitions\cite{Qi} can all be engineered into non-Abelian topological phases similar to a spinless $p+ip$ superconductor.  More recently, two groups\cite{1DwiresLutchyn,1DwiresOreg} recognized that topological superconductivity can be perhaps most easily engineered in \emph{one-dimensional} (1D) semiconducting wires deposited on an $s$-wave superconductor.  These proposals provide the first realistic experimental setting for Kitaev's\cite{1DwiresKitaev} 1D topological superconducting state.  Remarkably, the ends of such wires support a localized, zero-energy Majorana fermion\cite{1DwiresKitaev,1DwiresLutchyn,1DwiresOreg}.  Motivated by the exciting possibility of experimentally realizing this phase, we examine the prospect of exploiting 1D semiconducting wires for topological quantum computation.

The suitability of 1D wires for this purpose is by no means obvious.  One of the operations desired for topological quantum computation is braiding (though measurement-only approaches sidestep the need to physically braid particles\cite{MeasurementOnlyTQC}), and here non-Abelian statistics is critical.  While Majorana fermions can be transported, created, and fused in a physically transparent fashion by applying independently tunable gates to the wire, braiding and, in particular, non-Abelian statistics poses a serious puzzle.  Indeed, conventional wisdom holds that braiding statistics is ill-defined in 1D, since particles must pass through one another at some point during the exchange.  This problem can be surmounted by considering \emph{networks} of 1D wires, the simplest being the T-junction of Fig.\ \ref{Tjunction}.  Even in such networks, however, non-Abelian statistics does not immediately follow as recognized by Wimmer \emph{et al}.\cite{Wimmer}  For example, in a 2D $p+ip$ superconductor, vortices binding the Majoranas play an integral role in establishing non-Abelian statistics\cite{ReadGreen,Ivanov}.  We demonstrate that despite the absence of vortices in the wires, Majorana fermions in semiconducting wires exhibit non-Abelian statistics and transform under exchange exactly like vortices in a $p+ip$ superconductor.

We further propose experimental setups ranging from minimal circuits (involving one wire and a few gates) that can probe the Majorana fusion rules, to scalable networks that enable efficient exchange of many Majoranas.  
The `fractional Josephson effect'\cite{1DwiresKitaev,FuKane,MajoranaQSHedge,1DwiresLutchyn,1DwiresOreg}, along with Hassler \emph{et al.}'s recent proposal\cite{Hassler} enable readout of the topological qubits in this setting.  The relative ease with which 1D wires can be driven into a topological superconducting state, combined with the physical transparency of the manipulations, render the setups discussed here extremely promising venues for topological quantum information processing.  Although braiding of Majoranas alone does not permit universal quantum computation\cite{BravyiKitaev,UniversalTQC,TQCreview,Bonderson,BondersonPhaseGate}, implementation of the ideas introduced here would constitute a critical step towards this ultimate goal.

\section{Majorana fermions in `spinless' $p$-wave superconducting wires}

\begin{figure}
\centering{
  \includegraphics[width=3.2in]{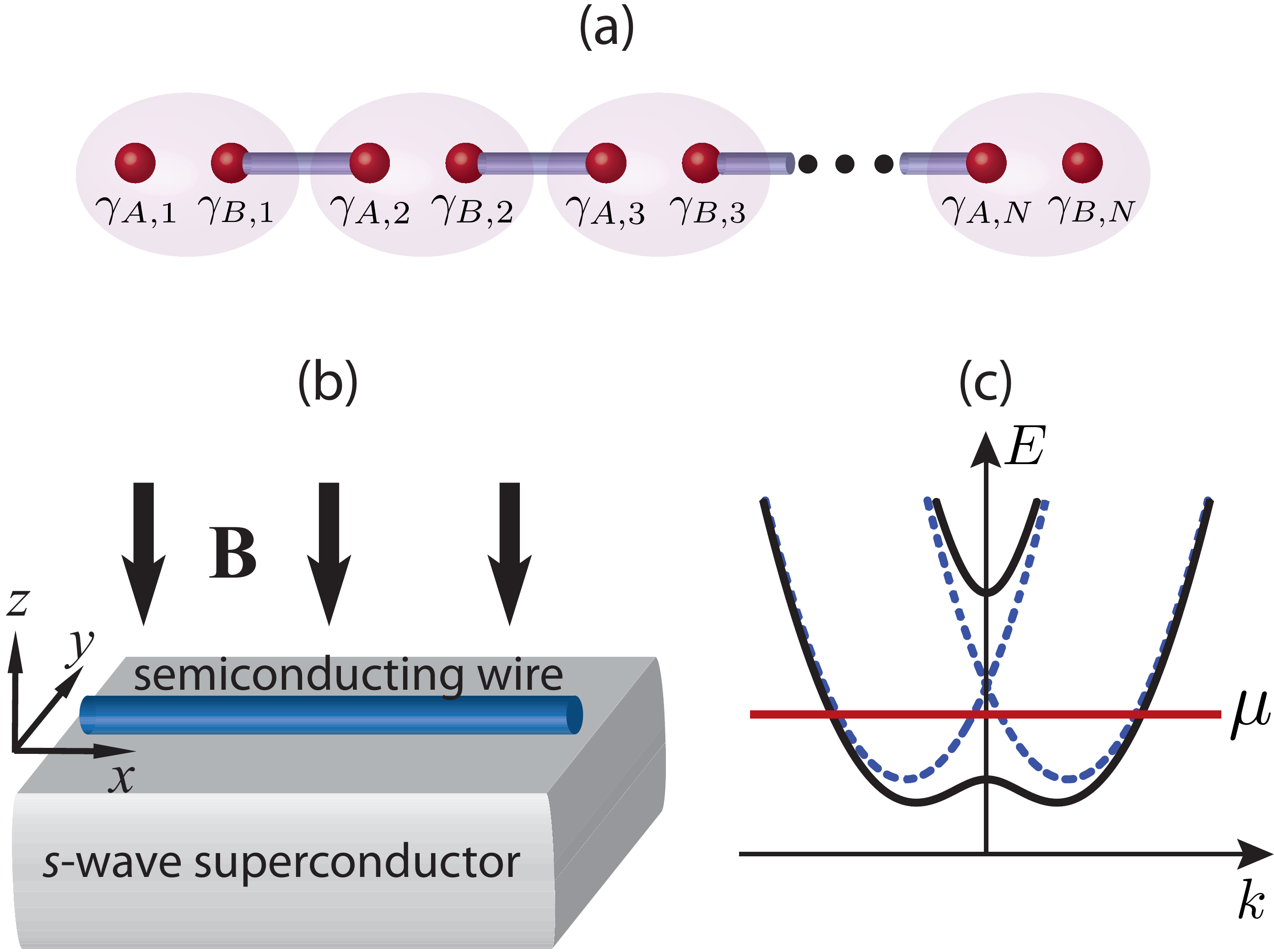}
  \caption{(a) Pictorial representation of the ground state of Eq.\ (\ref{Hkitaev}) in the limit $\mu = 0$, $t = |\Delta|$.  Each spinless fermion in the chain is decomposed in terms of two Majorana fermions $\gamma_{A,x}$ and $\gamma_{B,x}$.  Majoranas $\gamma_{B,x}$ and $\gamma_{A,x+1}$ combine to form an ordinary, finite energy fermion, leaving two zero-energy end Majoranas $\gamma_{A,1}$ and $\gamma_{B,N}$ as shown\cite{1DwiresKitaev}.  (b) A spin-orbit-coupled semiconducting wire deposited on an $s$-wave superconductor can be driven into a topological superconducting state exhibiting such end Majorana modes by applying an external magnetic field\cite{1DwiresLutchyn,1DwiresOreg}.  (c) Band structure of the semiconducting wire when ${\bf B} = 0$ (dashed lines) and ${\bf B} \neq 0$ (solid lines).  When $\mu$ lies in the band gap generated by the field, pairing inherited from the proximate superconductor drives the wire into the topological state.  }
  \label{MajoranaWire}}
\end{figure}

We begin by discussing the physics of a single wire.  Valuable intuition can be garnered from Kitaev's toy model for a spinless, $p$-wave superconducting $N$-site chain\cite{1DwiresKitaev}:
\begin{equation}
  H = -\mu \sum_{x = 1}^N c_x^\dagger c_x - \sum_{x = 1}^{N-1}(t c_x^\dagger c_{x+1} + |\Delta|e^{i\phi} c_x c_{x+1}+ h.c.)
  \label{Hkitaev}
\end{equation}
where $c_x$ is a spinless fermion operator and $\mu$, $t>0$, and $|\Delta|e^{i\phi}$ respectively denote the chemical potential, tunneling strength, and pairing potential.  The bulk- and end-state structure becomes particularly transparent in the special case\cite{1DwiresKitaev} $\mu = 0$, $t = |\Delta|$.
Here it is useful to express
\begin{equation}
  c_x = \frac{1}{2}e^{-i\frac{\phi}{2}}(\gamma_{B,x} + i \gamma_{A,x}),
  \label{MajoranaDecomposition}
\end{equation}
with $\gamma_{\alpha,x} = \gamma_{\alpha,x}^\dagger$ Majorana fermion operators satisfying $\{\gamma_{\alpha,x},\gamma_{\alpha',x'}\} = 2\delta_{\alpha\alpha'}\delta_{xx'}$.  These expressions expose the defining characteristics of Majorana fermions---they are their own antiparticle and constitute `half' of an ordinary fermion.  In this limit the Hamiltonian can be written as
\begin{equation}
  H = -it \sum_{x = 1}^{N-1}\gamma_{B,x}\gamma_{A,x+1}.
\end{equation}
Consequently, $\gamma_{B,x}$ and $\gamma_{A,x+1}$ combine to form an ordinary fermion $d_x = (\gamma_{A,x+1} + i\gamma_{B,x})/2$ which costs energy $2t$, reflecting the wire's bulk gap.  Conspicuously absent from $H$, however, are $\gamma_{A,1}$ and $\gamma_{B,N}$, which represent end-Majorana modes.  These can be combined into an ordinary (though highly non-local) zero-energy fermion $d_{\rm{end}} = (\gamma_{A,1} + i\gamma_{B,N})/2$.  Thus there are two degenerate ground states $|0\rangle$ and $|1\rangle = d^\dagger_{\rm end}|0\rangle$, where $d_{\rm end}|0\rangle = 0$, which serve as topologically protected qubit states.  Figure \ref{MajoranaWire}(a) illustrates this physics pictorially.

Away from this special limit the Majorana end states no longer retain this simple form, but survive provided the bulk gap remains finite\cite{1DwiresKitaev}.  This occurs when $|\mu| < 2t$, where a partially filled band pairs.  The bulk gap closes when $|\mu| = 2t$, and for larger $|\mu|$ a topologically trivial superconducting state without end Majoranas emerges.  Here pairing occurs in either a fully occupied or vacant band.

Realizing Kitaev's topological superconducting state experimentally requires a system which is effectively spinless---\emph{i.e.}, exhibits one set of Fermi points---and $p$-wave pairs at the Fermi energy.  Both criteria can be satisfied in a spin-orbit-coupled semiconducting wire deposited on an $s$-wave superconductor by applying a magnetic field\cite{1DwiresLutchyn,1DwiresOreg} [see Fig.\ \ref{MajoranaWire}(b)].  The simplest Hamiltonian describing such a wire reads
\begin{eqnarray}
  \mathcal{H} &=& \int dx\bigg{[} \psi^\dagger_x \bigg(-\frac{\hbar^2\partial_x^2}{2m}-\mu - i \hbar u \bold{\hat e}\cdot \bm{\sigma}\partial_x 
  \nonumber \\
  &-&\frac{g\mu_B B_z}{2} \sigma^z\bigg)\psi_x
  + (|\Delta| e^{i\varphi} \psi_{\downarrow x} \psi_{\uparrow x} + h.c.)\bigg{]}.
  \label{Hwire}
\end{eqnarray}
The operator $\psi_{\alpha x}$ corresponds to electrons with spin $\alpha$, effective mass $m$, and chemical potential $\mu$.  (We suppress the spin indices except in the pairing term.)  In the third term, $u$ denotes the (Dresselhaus\cite{Dresselhaus} and/or Rashba\cite{Rashba}) spin-orbit strength, and $\bm{\sigma} = (\sigma^x,\sigma^y,\sigma^z)$ is a vector of Pauli matrices.  This coupling favors aligning spins along or against the unit vector $\bf{\hat{e}}$, which we assume lies in the $(x,y)$ plane.  The fourth term represents the Zeeman coupling due to the magnetic field $B_z<0$.  Note that spin-orbit enhancement can lead to\cite{WinklerBook} $g\gg2$.  Finally, the last term reflects the spin-singlet pairing inherited from the $s$-wave superconductor via the proximity effect.

To understand the physics of Eq.\ (\ref{Hwire}), consider first $B_z = \Delta = 0$.  The dashed lines in Fig.\ \ref{MajoranaWire}(c) illustrate the band structure here---clearly no `spinless' regime is possible.  Introducing a magnetic field generates a band gap $\propto |B_z|$ at zero momentum as the solid line in Fig.\ \ref{MajoranaWire}(c) depicts.  When $\mu$ lies inside of this gap the system exhibits only a single pair of Fermi points as desired.  Turning on $\Delta$ which is weak compared to the gap then effectively $p$-wave pairs fermions in the lower band with momentum $k$ and $-k$, driving the wire into Kitaev's topological phase\cite{1DwiresLutchyn,1DwiresOreg}.  [Singlet pairing in Eq.\ (\ref{Hwire}) generates $p$-wave pairing because spin-orbit coupling favors opposite spins for $k$ and $-k$ states in the lower band.]  Quantitatively, realizing the topological phase requires\cite{1DwiresLutchyn,1DwiresOreg} $|\Delta| < g\mu_B |B_z|/2$, which we hereafter assume holds.  The opposite limit $|\Delta|>g\mu_B |B_z|/2$ effectively violates the `spinless' criterion since pairing strongly intermixes states from the upper band, producing an ordinary superconductor without Majorana modes.

In the topological phase, the connection to Eq.\ (\ref{Hkitaev}) becomes more explicit when $g\mu_B |B_z| \gg m u^2, |\Delta|$ where the spins nearly polarize.  One can then project Eq.\ (\ref{Hwire}) onto a simpler one-band problem by writing $\psi_{\uparrow x}\sim \frac{u(e_y + i e_x)}{g\mu_B |B_z|}\partial_x\Psi_x$ and $\psi_{\downarrow x}\sim \Psi_x$, with $\Psi_x$ the lower-band fermion operator.  To leading order, one obtains
\begin{eqnarray}
  \mathcal{H}_{\rm eff} &\sim& \int dx \bigg{[}\Psi^\dagger_x\left(-\frac{\hbar^2\partial_x^2}{2m}-\mu_{\rm{eff}}\right)\Psi_x
  \nonumber \\
  &+& \left(|\Delta_{\rm{eff}}|e^{i\varphi_{\rm{eff}}}\Psi_x \partial_x \Psi_x + h.c. \right)\bigg{]},
  \label{Heff}
\end{eqnarray}
where $\mu_{\rm{eff}} = \mu+ g\mu_B|B_z|/2$ and the effective $p$-wave pair field reads
\begin{equation}
  |\Delta_{\rm{eff}}|e^{i\varphi_{\rm{eff}}} \approx \frac{u|\Delta|}{g\mu_B|B_z|}e^{i\varphi}(e_y+i e_x).
  \label{DeltaEff}
\end{equation}
The dependence of $\varphi_{\rm{eff}}$ on ${\bf \hat{e}}$ will be important below when we consider networks of wires.  Equation (\ref{Heff}) constitutes an effective low-energy Hamiltonian for Kitaev's model in Eq.\ (\ref{Hkitaev}) in the low-density limit.  From this perspective, the existence of end-Majoranas in the semiconducting wire becomes manifest.  We exploit this correspondence below when addressing universal properties such as braiding statistics, which must be shared by the topological phases described by Eq.\ (\ref{Hwire}) and the simpler lattice model, Eq.\ (\ref{Hkitaev}).

We now seek a practical method to manipulate Majorana fermions in the wire.  As motivation, consider applying a gate voltage to adjust $\mu$ uniformly across the wire.  The excitation gap obtained from Eq.\ (\ref{Hwire}) at $k = 0$ varies with $\mu$ via
\begin{equation}
  E_{\rm{gap}}(k = 0) = \left|\frac{g\mu_B |B_z|}{2}-\sqrt{|\Delta|^2 + \mu^2}\right|.
\end{equation}
For $|\mu|<\mu_c = \sqrt{(g\mu_B B_z/2)^2 -|\Delta|^2}$ the topological phase with end Majoranas emerges, while for $|\mu|>\mu_c$ a topologically trivial phase appears.  Applying a gate voltage uniformly thus allows one to create or remove the Majorana fermions.  However, when $|\mu| = \mu_c$ the bulk gap closes, and the excitation spectrum at small momentum behaves as $E_{\rm{gap}}(k) \approx \hbar v|k|$, with velocity $v = 2u |\Delta|/(g\mu_B |B_z|)$.  The gap closure is clearly undesirable, since we would like to manipulate Majorana fermions without generating additional quasiparticles.

\begin{figure}
\centering{
  \includegraphics[width=2.5in]{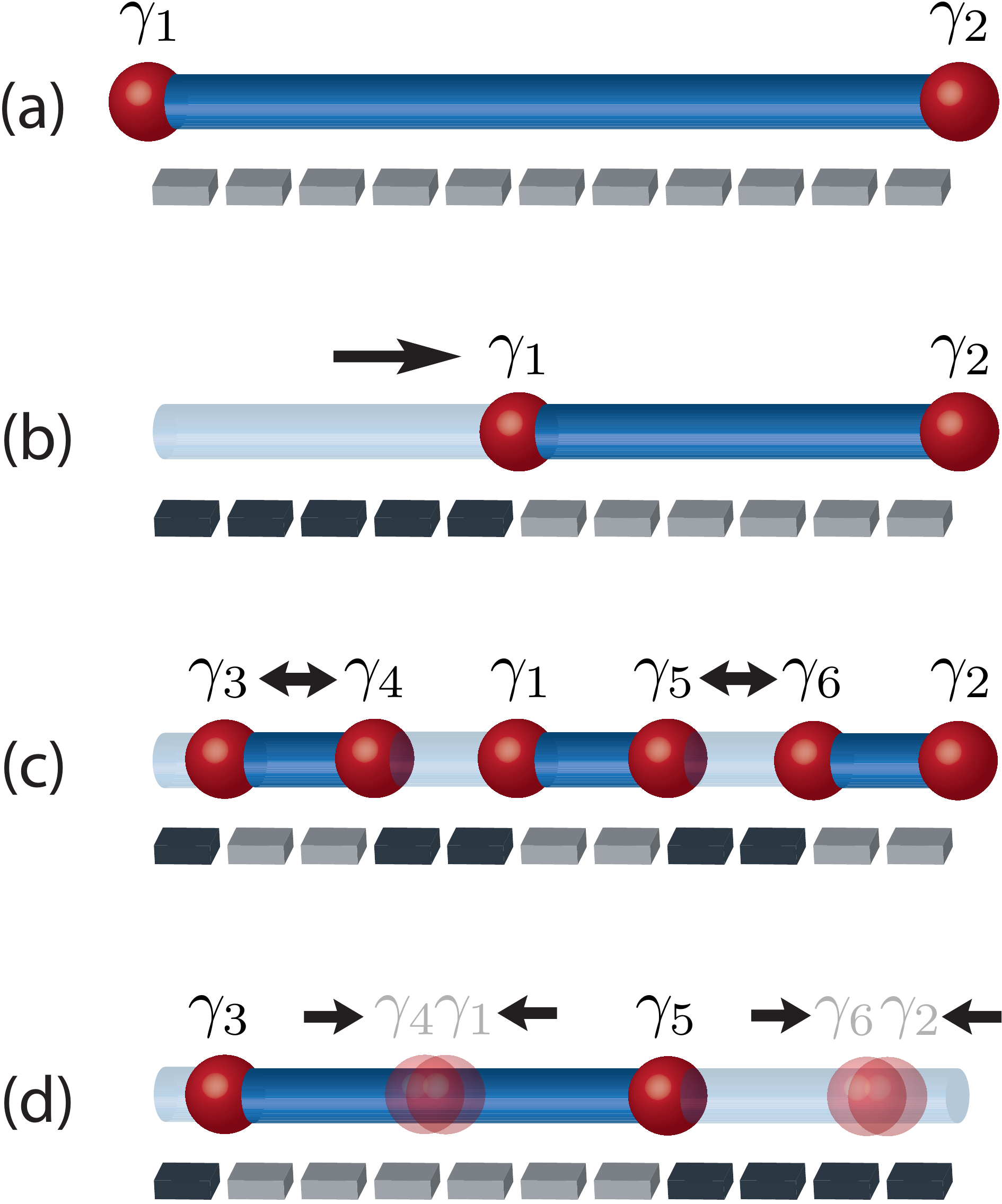}
  \caption{Applying a `keyboard' of individually tunable gates to the wire allows one to locally control which regions are topological (dark blue) and non-topological (light blue), and hence manipulate Majorana fermions while maintaining the bulk gap.  As (a) and (b) illustrate, sequentially applying the leftmost gates drives the left end of the wire non-topological, thereby transporting $\gamma_1$ rightward.  Nucleating a topological section of the wire from an ordinary region or vice versa creates pairs of Majorana fermions out of the vacuum as in (c).  
Similarly, removing a topological region entirely or connecting two topological regions as sketched in (d) fuses Majorana fermions into either the vacuum or a finite-energy quasiparticle.} 
  \label{MajoranaManipulation}}
\end{figure}

This problem can be circumvented by employing a `keyboard' of locally tunable gates as shown in Fig.\ \ref{MajoranaManipulation}, each of which impacts $\mu$ over a finite length $L_{\rm gate}$ of the wire.  When a given gate locally tunes the chemical potential across $|\mu| = \mu_c$, a finite excitation gap $E_{\rm{gap}} \sim \hbar v \pi/L_{\rm gate}$ remains.  (Roughly, the gate creates a potential well that supports only $k$ larger than $\sim \pi/L_{\rm gate}$.)  Assuming $g\mu_B |B_z|/2 \sim 2|\Delta|$ and $\hbar u \sim 0.1$eV\AA~yields a velocity $v \sim 10^4$m/s; the gap for a 0.1$\mu$m wide gate is then of order 1K.  We consider this a conservative estimate---heavy-element wires such as InSb and/or narrower gates could generate substantially larger gaps.  

\begin{figure}
\centering{
  \includegraphics[width=3.2in]{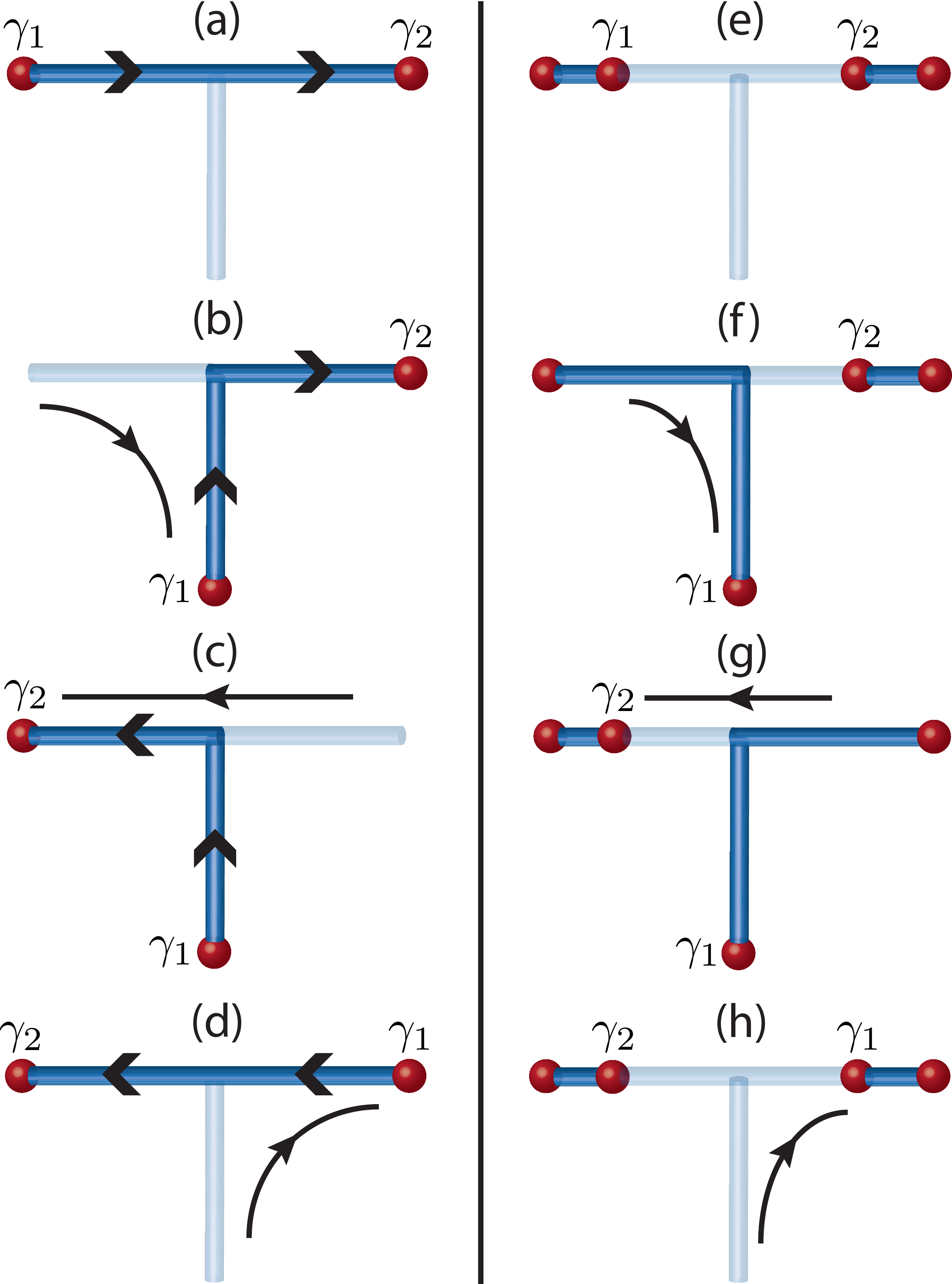}
  \caption{A T-junction allows for adiabatic exchange of two Majorana fermions bridged by either a topological region (dark blue lines) as in (a)-(d), or a non-topological region (light blue lines) as in (e)-(h).  Transport of Majorana fermions is achieved by gates as outlined in Fig.\ \ref{MajoranaManipulation}.  The arrows along the topological regions in (a)-(d) are useful for understanding the non-Abelian statistics as outlined in the main text.}
  \label{Tjunction}}
\end{figure}

Local gates allow Majorana fermions to be transported, created, and fused as outlined in Fig.\ \ref{MajoranaManipulation}.  As one germinates pairs of Majorana fermions, the ground state degeneracy increases as does our capacity to topologically store quantum information in the wire.  Specifically, $2n$ Majoranas generate $n$ ordinary zero-energy fermions whose occupation numbers specify topological qubit states.  Adiabatically braiding the Majorana fermions would enable manipulation of the qubits, but is not possible in a single wire.  Thus we now turn to the simplest arrangement which allows for exchange---the T-junction of Fig.\ \ref{Tjunction}.

\section{Majorana Braiding and non-Abelian statistics} 

First, we explore the physics at the junction where the wires in Fig.\ \ref{Tjunction} meet (see the Supplementary Material for a more detailed discussion).  It will be useful to view the T-junction as composed of three segments whose ends meet at a point.  When only one segment realizes a topological phase, a single zero-energy Majorana fermion exists at the junction.  When two topological segments meet at the junction, as in Figs.\ \ref{Tjunction}(a) and (b), generically no Majorana modes exist there.  To see this, imagine decoupling the two topological segments so that two Majorana modes in close proximity exist at the junction; restoring the coupling generically combines these Majoranas into an ordinary, finite-energy fermion.

As an illustrative example, consider the setup of Fig.\ \ref{Tjunction}(a) and model the left and right topological segments by Kitaev's model with $\mu = 0$ and $t = |\Delta|$ in Eq.\ (\ref{Hkitaev}).  [For simplicity we will exclude the non-topological vertical wire in Fig.\ \ref{Tjunction}(a).]  Suppose furthermore that the $\phi = \phi_{L/R}$ in the left/right chains and that the fermion $c_{L,N}$ at site $N$ of the left chain couples weakly to the fermion $c_{R,1}$ at site 1 of the right chain via $H_\Gamma = -\Gamma(c^{\dagger}_{L,N} c_{R,1} + h.c.)$.  Using Eq.\ (\ref{MajoranaDecomposition}), the end Majoranas at the junction couple as follows,
\begin{equation}
  H_\Gamma \sim -\frac{i \Gamma}{2}\cos\left(\frac{\phi_L-\phi_R}{2}\right)\gamma_{B,N}^{L}\gamma_{A,1}^{R}
  \label{EndMajoranaCoupling}
\end{equation}
and therefore generally combine into an ordinary fermion\cite{1DwiresKitaev}.
An exception occurs when the regions form a $\pi$-junction---that is, when $\phi_L-\phi_R = \pi$---which fine-tunes their coupling to zero.  Importantly, coupling between end Majoranas in the semiconductor context is governed by the same $\phi_L-\phi_R$ dependence as in Eq.\ (\ref{EndMajoranaCoupling})\cite{1DwiresLutchyn,1DwiresOreg}.

Finally, when all three segments are topological, again only a single Majorana mode exists at the junction without fine-tuning.  Three Majorana modes appear only when all pairs of wires simultaneously form mutual $\pi$ junctions (which is possible as described in the Supplementary Material, since the superconducting phases are defined with respect to a direction in each wire).  Recall from Eq.\ (\ref{DeltaEff}) that the spin-orientation favored by spin-orbit coupling determines the effective superconducting phase of the semiconducting wires.  Two wires at right angles to one another therefore exhibit a $\pi/2$ phase difference, well away from the pathological limits mentioned above.

The T-junction permits exchange of Majoranas from either the same or different topological regions.  First, consider the configuration of Fig.\ \ref{Tjunction}(a) where the horizontal wire resides in a topological phase while the vertical wire is non-topological.  Counterclockwise exchange of $\gamma_1$ and $\gamma_2$ can be implemented as outlined in Figs.\ \ref{Tjunction}(b)-(d).  Here, one shuttles $\gamma_1$ to the junction by making the left end non-topological; transports $\gamma_1$ downward by driving the vertical wire into a topological phase; transports $\gamma_2$ leftward in a similar fashion; and finally directs $\gamma_1$ up and to the right.  Exchange of two Majorana fermions connected by a non-topological region as in Fig.\ \ref{Tjunction}(e) can be similarly achieved---counterclockwise exchange of $\gamma_1$ and $\gamma_2$ proceeds as sketched in Figs.\ \ref{Tjunction}(f)-(h).

While the Majoranas can now be exchanged, non-Abelian statistics is not obvious in this context.  Recall how non-Abelian statistics of vortices arises in a spinless 2D $p+ip$ superconductor\cite{ReadGreen,Ivanov}, following Ivanov's approach.  Ultimately, this can be deduced by considering two vortices which bind Majorana fermions $\gamma_1$ and $\gamma_2$.  Since the spinless fermion operators effectively change sign upon advancing the superconducting phase by $2\pi$, one introduces branch cuts emanating from the vortices; crucially, a Majorana fermion changes sign whenever crossing such a cut.  Upon exchanging the vortices, $\gamma_2$ (say) crosses the branch cut emanating from the other vortex, leading to the transformation rule $\gamma_1 \rightarrow \gamma_2$ and $\gamma_2\rightarrow-\gamma_1$ which is generated by the unitary operator $U_{12} = \exp(\pi\gamma_2\gamma_1/4)$.  With many vortices, the analogous unitary operators $U_{ij}$ corresponding to the exchange of $\gamma_i$ and $\gamma_j$ do not generally commute, implying non-Abelian statistics.

Following an approach similar to that of Stern \emph{et al}.\cite{SternNonAbelian}, we now argue that Majorana fermions in semiconducting wires transform exactly like those bound to vortices under exchange, and hence also exhibit non-Abelian statistics.  This can be established most simply by considering the exchange of two Majorana fermions $\gamma_1$ and $\gamma_2$ as illustrated in Figs.\ \ref{Tjunction}(a)-(d).  At each step of the exchange, there are two degenerate ground states $|0\rangle$ and $|1\rangle = f^\dagger |0\rangle$, where $f = (\gamma_1 + i\gamma_2)/2$ annihilates $|0\rangle$.  In principle, one can deduce the transformation rule from the Berry phases $\chi_n \equiv {\rm Im}\int dt \langle n|\partial_t|n\rangle$ acquired by the many-body ground states $|n\rangle = |0\rangle$ and $|1\rangle$, though in practice these are hard to evaluate.

Since exchange statistics is a universal property, however, we are free to deform the problem to our convenience provided the energy gap remains finite.   As a first simplification, since the semiconductor Hamiltonian and Kitaev's model in Eq.\ (\ref{Hkitaev}) can be smoothly connected, let us consider the case where each wire in the T-junction is described by the latter.  More importantly, we further deform Kitaev's Hamiltonian to be \emph{purely real} as we exchange $\gamma_{1,2}$.  The states $|0\rangle$ and $|1\rangle$ can then also be chosen real, leading to an enormous simplification: while these states still evolve nontrivially \emph{the Berry phase accumulated during this evolution vanishes}.

For concreteness, we deform the Hamiltonian such that $\mu<0$ and $t = \Delta = 0$ in the non-topological regions of Fig.\ \ref{Tjunction}.  For the topological segments, reality implies that the superconducting phases must be either 0 or $\pi$.  It is useful to visualize the sign choice for the pairing with arrows as in Fig.\ \ref{Tjunction}.  (To be concrete, we take the pairing $|\Delta|e^{i\phi} c_j c_{j+1}$ such that the site indices increase moving rightward/upward in the horizontal/vertical wires; the case $\phi = 0$ then corresponds to rightward/upward arrows, while leftward/downward arrows indicate $\phi = \pi$.)  To avoid generating $\pi$ junctions, when two topological segments meet at the junction, one arrow must point into the junction while the other must point out.  With this simple rule in mind, we see in Fig.\ \ref{Tjunction} that although we can successfully swap the Majoranas while keeping the Hamiltonian real, we inevitably end up reversing the arrows along the topological region.  In other words, the sign of the pairing has flipped relative to our initial Hamiltonian.

To complete the exchange then, we must perform a gauge transformation which restores the Hamiltonian to its original form.  This can be accomplished by multiplying all fermion creation operators by $i$; in particular, $f^\dagger = (\gamma_1-i\gamma_2)/2 \rightarrow i f^\dagger = (\gamma_2+i \gamma_1)/2$.  It follows that $\gamma_1 \rightarrow \gamma_2$ and $\gamma_2\rightarrow -\gamma_1$, which the unitary transformation $U_{12} = \exp(\pi\gamma_2\gamma_1/4)$ generates as in the 2D $p+ip$ case.  We stress that this result applies also in the physically relevant case where gates transport the Majoranas while the superconducting phases remain fixed; we have merely used our freedom to deform the Hamiltonian to expose the answer with minimal formalism.  Additionally, since Figs.\ \ref{Tjunction}(e)-(h) also represent a counterclockwise exchange, it is natural to expect the same result for this case.  The Supplementary Material analyzes both types of exchanges from a complementary perspective (and when the superconducting phases are held fixed), confirming their equivalence.  There we also establish rigorously that in networks supporting arbitrarily many Majoranas exchange is implemented by a set of unitary operators $U_{ij}$ analogous to those in a 2D $p+ip$ superconductor.  Thus the statistics is non-Abelian as advertised.

\section{Discussion}

\begin{figure}
\centering{
  \includegraphics[width=3.2in]{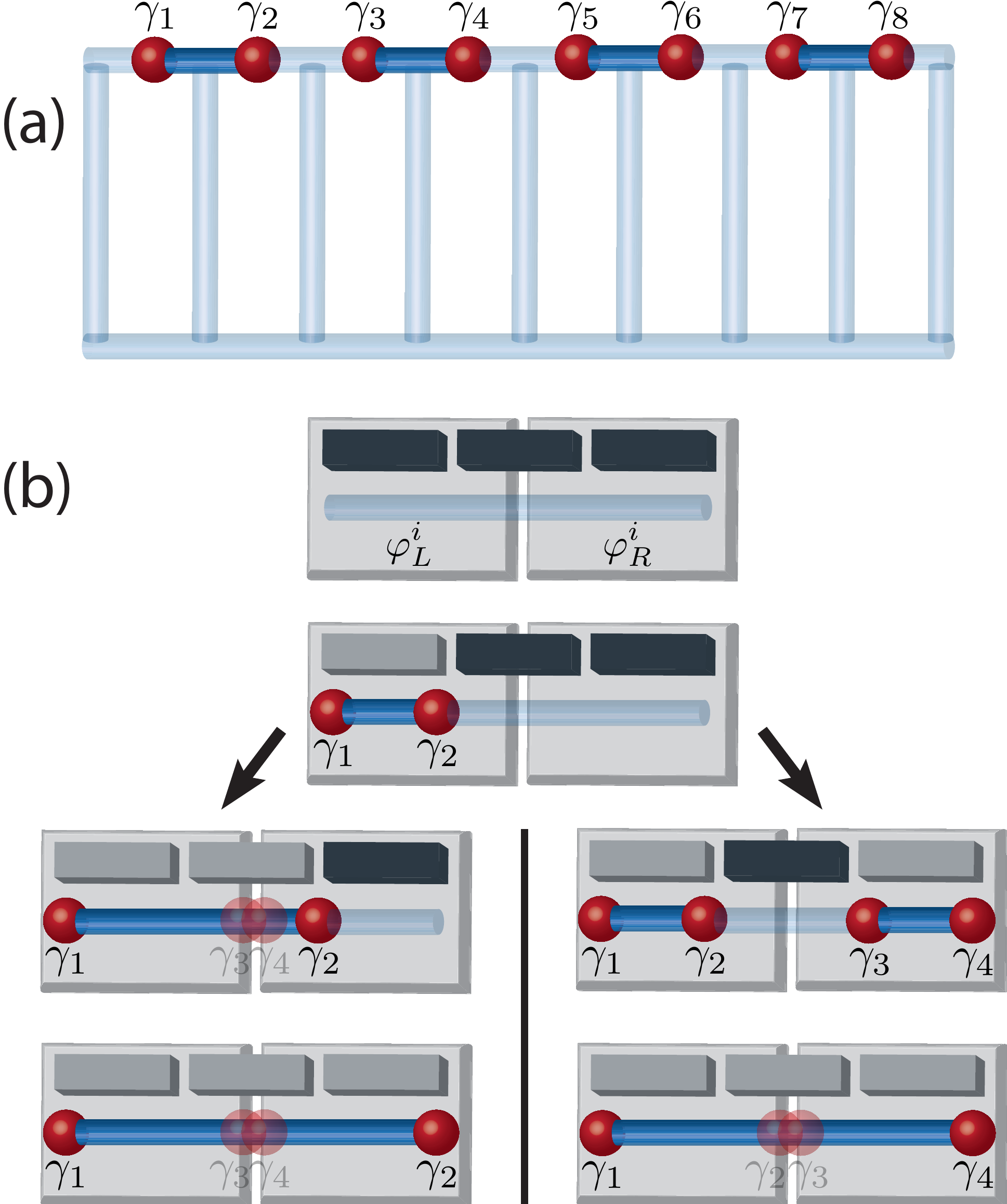}
  \caption{(a) Example of a semiconductor wire network which allows for efficient exchange of many Majorana fermions.  Adjacent Majoranas can be exchanged as in Fig.\ \ref{Tjunction}, while non-adjacent Majoranas can be transported to the lower wire to be similarly exchanged.  (b) Minimal setup designed to detect the non-trivial Majorana fusion rules.  Majoranas $\gamma_{1,2}$ are first created out of the vacuum.  In the left path, $\gamma_2$ is shuttled rightward, and Majoranas $\gamma_{3,4}$ always combine to form a finite-energy state which is unoccupied.  In the right path, $\gamma_{3,4}$ are also created out of the vacuum, and then $\gamma_2$ and $\gamma_3$ fuse with 50\% probability into either the vacuum or a finite-energy quasiparticle.  The Josephson current flowing across the junction allows one to deduce the presence or absence of this extra quasiparticle.  }
  \label{ExptSetups}}
\end{figure}

The keyboard of gates shown in Fig.\ \ref{MajoranaManipulation} and the T-junction of Fig.\ \ref{Tjunction} provide the basic elements allowing manipulation of topological qubits in semiconducting wires.  In principle, a single T-junction can support numerous well-separated Majorana modes, each of which can be exchanged with any other.  (First, create many Majoranas in the horizontal wire of the T-junction.  To exchange a given pair, shuttle all intervening Majoranas down to the end of the vertical wire and then carry out the exchange using the methods of Fig.\ \ref{Tjunction}.)  However, networks consisting of several T-junctions---such as the setup of Fig.\ \ref{ExptSetups}(a)---enable more efficient Majorana exchange.  In the figure, all adjacent Majorana fermions can be immediately swapped using Fig.\ \ref{Tjunction}, while non-adjacent Majoranas can be shuttled down to the lower wire to be exchanged.  This `ladder' configuration straightforwardly scales up by introducing additional `rungs' and/or `legs'.  

As Fu and Kane suggested in the topological insulator context\cite{FuKane}, fusing Majorana fermions across a Josephson junction provides a readout method for the topological qubit states.  We illustrate the physics with the schematic setup of Fig.\ \ref{ExptSetups}(b), which extends the experiments proposed in Refs.\ \onlinecite{1DwiresLutchyn,1DwiresOreg} to allow the Majorana fusion rules to be directly probed.  Here a semiconducting wire bridges two $s$-wave superconductors with initial phases $\varphi^{i}_{L/R}$; we assume $\Delta\varphi^i \equiv \varphi^i_L-\varphi^i_R \neq \pi$.  Three gates drive the wire from an initially non-topological ground state into a topological phase.  Importantly, the order in which one applies these gates qualitatively affects the physics.  As we now discuss, only in the left path of Fig.\ \ref{ExptSetups}(b) can the qubit state at the junction be determined in a single measurement.  

Consider first germinating Majorana fermions $\gamma_1$ and $\gamma_2$ by applying the left gate.  Assuming $f_A = (\gamma_1 + i\gamma_2)/2$ initially costs finite energy as $\gamma_1$ and $\gamma_2$ separate, the system initializes into a ground state with $f_A$ unoccupied.  Applying the central and then right gates shuttles $\gamma_2$ to the other end [see the left path of Fig.\ \ref{ExptSetups}(b)].  Since a narrow insulating barrier separates the superconductors, an ordinary fermion $f_B = (\gamma_3 + i\gamma_4)/2$ arises from two coupled Majoranas $\gamma_{3,4}$ at the junction.  Similar to Eq.\ (\ref{EndMajoranaCoupling}), the energy of this mode is well-captured by\cite{1DwiresKitaev,1DwiresLutchyn,1DwiresOreg} $H_J \sim i\epsilon^i\gamma_3 \gamma_4 = \epsilon^i(2f_B^\dagger f_B-1)$ where $\epsilon^i = \delta \cos(\Delta\varphi^i/2)$ with non-universal $\delta$.  The system has been prepared in a ground state, so the $f_B$ fermion will be absent if $\epsilon^i>0$ but occupied otherwise.  

Suppose we now vary the phase difference across the junction away from its initial value to $\Delta\varphi$.  The measured Josephson current (see Supplementary Material for a pedagogical derivation) will then be
\begin{eqnarray}
  I &=& \frac{2e}{\hbar}\frac{d E}{d\Delta\varphi} = \frac{e\delta}{\hbar}{\rm sgn}(\epsilon^i)\sin(\Delta\varphi/2) + I_{2e},
  \label{JosephsonCurrent}
\end{eqnarray}
where $E$ is the ground-state energy and $I_{2e}$ denotes the usual Cooper-pair-tunneling contribution.  The first term on the right reflects single-electron tunneling originating from the Majoranas $\gamma_{3,4}$.  This `fractional' Josephson current exhibits $4\pi$ periodicity in $\Delta\varphi$, but $2\pi$ periodicity in the initial phase difference $\Delta\varphi^i$.  

The right path in Fig.\ \ref{ExptSetups}(b) yields very different results, reflecting the nontrivial Majorana fusion rules.  Here, after creating $\gamma_{1,2}$ one applies the rightmost gate to nucleate another pair $\gamma_{3,4}$.  Assuming $f_A$ and $f_B$ defined as above initially cost finite energy, the system initializes into the ground state $|0,0\rangle$ satisfying $f_{A/B}|0,0\rangle = 0$.  Applying the central gate then fuses $\gamma_2$ and $\gamma_3$ at the junction.  To understand the outcome, it is useful to re-express the ground state in terms of $f_A' = (\gamma_1+i\gamma_4)/2$ and $f_B' = (\gamma_2+i\gamma_3)/2$.  In this basis $|0,0\rangle = (|0',0'\rangle-i|1',1'\rangle)/\sqrt{2}$, where $f_{A,B}'$ annihilate $|0',0'\rangle$ and $|1',1'\rangle = f_A'^\dagger f_B'^\dagger|0',0'\rangle$.  Following our previous discussion, $f_B'$ acquires finite energy at the junction, lifting the degeneracy between $|0',0'\rangle$ and $|1',1'\rangle$.  Measuring the Josephson current then collapses the wavefunction with 50\% probability onto either the ground state, or an excited state with an extra quasiparticle localized at the junction.  In the former case Eq.\ (\ref{JosephsonCurrent}) again describes the current, while in the latter the fractional contribution simply changes sign.  

In more complex networks such as that of Fig.\ \ref{ExptSetups}(a), fusing the Majoranas across a Josephson junction---and in particular measuring the sign of the fractional Josephson current---similarly allows qubit readout.  Alternatively, the interesting recent proposal of Hassler \emph{et al}.\cite{Hassler} for reading qubit states via ancillary non-topological flux qubits can be adapted to these setups (and indeed was originally discussed in terms of an isolated semiconducting wire\cite{Hassler}).  

To conclude, we have introduced a surprising new venue for braiding, non-Abelian statistics, and topological quantum information processing---networks of \emph{one-dimensional} semiconducting wires.  From a fundamental standpoint, the ability to realize non-Abelian statistics in this setting is remarkable.  Perhaps even more appealing, however, are our proposal's physical transparency and experimental promise, particularly given the feats already achieved in Ref.\ \onlinecite{nanowireExpt}.  While topological quantum information processing in wire networks requires much experimental progress, observing the distinct fusion channels characteristic of the two paths of Fig.\ \ref{ExptSetups}(b) would provide a remarkable step en route to this goal.  And ultimately, if braiding in this setting can be supplemented by a $\pi/8$ phase gate and topological charge measurement of four Majoranas, wire networks may provide a feasible path to \emph{universal} quantum computation\cite{BravyiKitaev,UniversalTQC,TQCreview,Bonderson,BondersonPhaseGate}.

\section{Supplementary Material}

\subsection{Properties of the T-junction}

\begin{figure}
\centering{
  \includegraphics[width=3in]{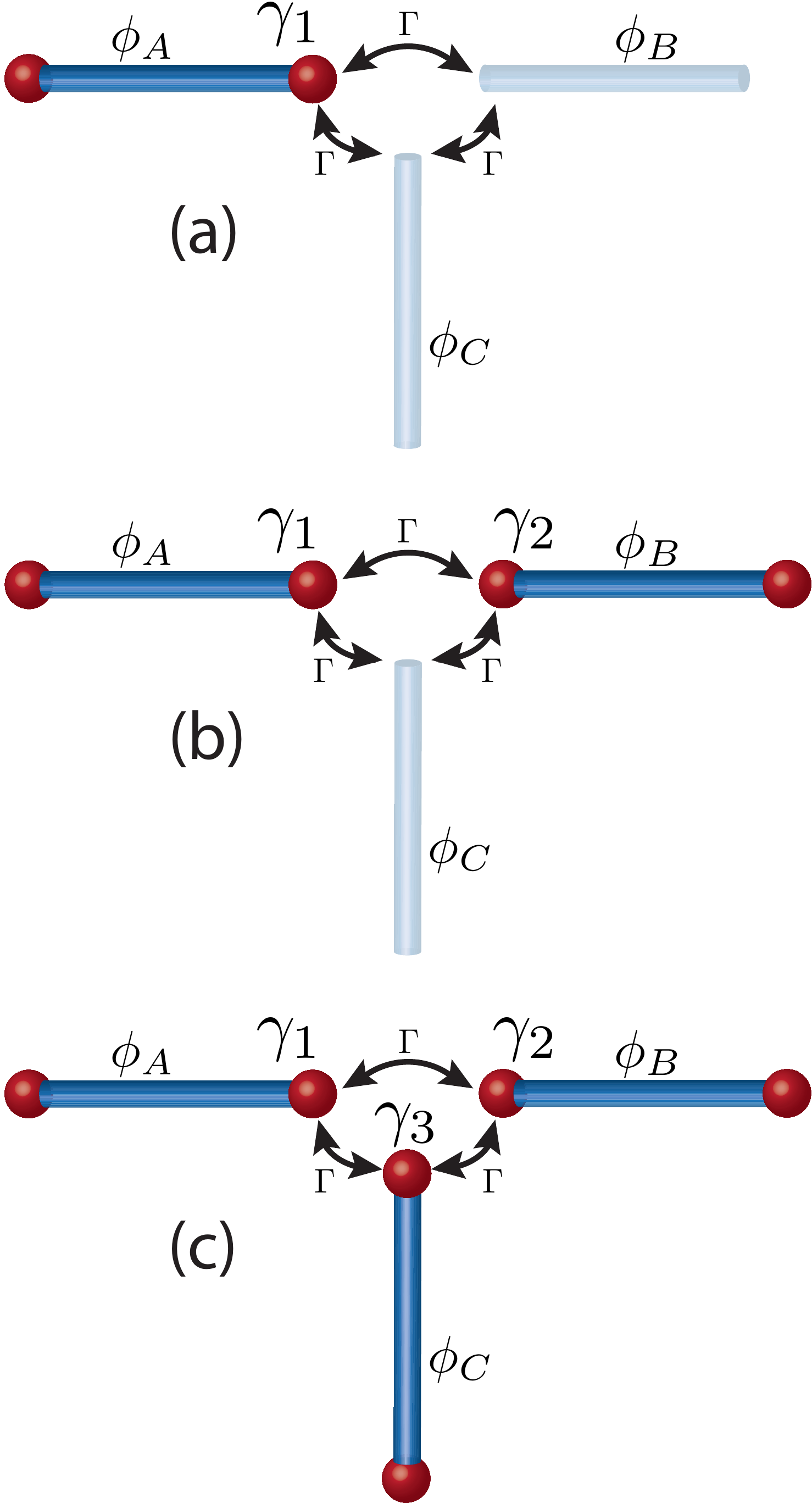}
  \caption{T-junction viewed as three wire segments with $p$-wave superconducting phases $\phi_{A,B,C}$.  The ends of each segment are coupled via tunneling with amplitude $\Gamma$ as shown.  (a) When only one segment is topological, the tunneling can not destroy the Majorana $\gamma_1$ at the junction.  (b) Two topological regions meeting at the junction leads to the end Majoranas $\gamma_1$ and $\gamma_2$ generally combining into an ordinary, finite-energy fermion, unless the topological wires form a $\pi$ junction.  (c) When all the three wires are topological, the Majoranas $\gamma_{1,2,3}$ generally combine to form a finite-energy fermion and a single topologically protected Majorana.  All three Majoranas remain at zero energy only when all three wire segments form mutual $\pi$ junctions.  }
  \label{TjunctionPropertiesFig}}
\end{figure}

Here we investigate in greater detail the properties of the junction in Fig.\ \ref{Tjunction} where the three wire segments meet.  There are three cases to consider, corresponding to the situations where one, two, or all three of the wire segments emanating from the junction reside in a topological superconducting state.  It is conceptually simplest to address each case by viewing the T-junction as composed of three independent wire segments as in Fig.\ \ref{TjunctionPropertiesFig}, which initially decouple from one another.  In this limit a single Majorana exists at the end of each topological segment.  One can then straightforwardly couple the wire segments at the junction and explore the fate of the Majorana end states.

Suppose that the phases of the $p$-wave pair fields in each region are $\phi_{A/B/C}$ as shown in Fig.\ \ref{TjunctionPropertiesFig} [in the semiconductor wire context, these phases correspond to $\varphi_{\rm eff}$ in Eq.\ (6) of the main text].  To be precise, if the wires are described by a lattice model, we define these phases relative to a pairing term $|\Delta|e^{i\phi_\alpha}c_j c_{j+1}$ such that the site indices increase moving rightward in the horizontal wires and upward in the vertical wire.  A similar convention can be employed in the semiconductor wire context.  Now suppose we allow single-electron tunneling between the ends of each segment, with amplitude $\Gamma$ as shown schematically in Fig.\ \ref{TjunctionPropertiesFig}.  (Pairing between electrons residing at the ends of each region is also generally allowed, but does not change any qualitative results below and will therefore be neglected.)  For convenience we will assume that the tunneling strength is weak compared to the bulk gaps in the wires, which will allow us to focus solely on the Majorana end states; our conclusions, however, are more general and do not require this assumption.

In the setup of Fig.\ \ref{TjunctionPropertiesFig}(a) with only one topological region, the Majorana $\gamma_1$ is qualitatively unaffected by the coupling to the non-topological wires.  At most its wavefunction can be quantitatively modified, but it necessarily remains at zero energy.  This reflects the familiar topological protection of a single isolated Majorana mode in a gapped system.  

With two of the three wires topological as in Fig.\ \ref{TjunctionPropertiesFig}(b), the end Majoranas $\gamma_{1}$ and $\gamma_2$ generally combine into an ordinary finite-energy fermion, except with fine-tuning.  To a good approximation, the Majoranas couple through a Hamiltonian\cite{1DwiresKitaev,1DwiresLutchyn,1DwiresOreg}
\begin{equation}
  H_{12} \propto -i\Gamma \cos\left(\frac{\phi_A-\phi_B}{2}\right)\gamma_1\gamma_2.
  \label{H12}
\end{equation}
This was discussed in the main text in the context of two wires described by Kitaev's toy model in a particular limit, but is qualitatively rather general---the $4\pi$ periodicity in $\phi_A-\phi_B$ has a topological origin\cite{1DwiresKitaev}.  For instance, end Majoranas in two topological semiconducting wires coupled through an ordinary region exhibit the same phase dependence as above\cite{1DwiresLutchyn,1DwiresOreg}.  Equation (\ref{H12}) demonstrates that $\gamma_1$ and $\gamma_2$ remain zero-energy modes only when the topological wires form a $\pi$ junction, \emph{i.e.}, when $\phi_A = \phi_B+\pi$.  

Finally, consider the case shown in Fig.\ \ref{TjunctionPropertiesFig}(c) where all three segments are topological.  Here the Majoranas $\gamma_{1,2,3}$ couple via
\begin{eqnarray}
  H_{123} &\propto& -i\Gamma \bigg{[}\cos\left(\frac{\phi_A-\phi_B}{2}\right)\gamma_1\gamma_2
  \nonumber \\
  &+& \cos\left(\frac{\phi_C-\phi_B}{2}\right)\gamma_3\gamma_2 + \sin\left(\frac{\phi_A-\phi_C}{2}\right)\gamma_1\gamma_3\bigg{]}.
  \nonumber \\
  \label{Trijunction}
\end{eqnarray}
Note the sine function determining the coupling between $\gamma_1$ and $\gamma_3$, which arises because of the conventions we chose for defining $\phi_\alpha$ above.  Recall that to make the problem well-defined, we needed to define the phases with respect to a particular direction in each wire; otherwise there is an ambiguity of $\pi$ in the definition, since for instance $|\Delta|e^{i\phi} c_j c_{j+1} = |\Delta|e^{i(\phi+\pi)}c_{j+1}c_j$.  We defined the phases such that the site indices increase upon moving rightward or upward in the wires. But this implies that the site indices in both the left and bottom wires increase upon moving towards the junction, in contrast to all other pairs of wires.  It follows that
the splitting of $\gamma_1$ and $\gamma_3$ is proportional to $\cos[(\phi_A-\phi_C-\pi)/2] = \sin[(\phi_A-\phi_C)/2]$.  Hence with our conventions a $\pi$ junction between these two regions actually corresponds to the case $\phi_A = \phi_C$.  

The Hamiltonian $H_{123}$ implies that $\gamma_{1,2,3}$ all remain zero-energy modes only when $\phi_A = \phi_C = \phi_B+\pi$, where all pairs of wires form mutual $\pi$ junctions.  (This remains true even when coupling to the ordinary gapped states is included.)  Aside from this fine-tuned limit, however, $H_{123}$ always supports one zero-energy Majorana mode and one ordinary finite-energy fermion.  As an illustration, consider the case $\phi_A = \phi_B+\pi$ and $\phi_C\neq \phi_A$, so that only the horizontal wires form a $\pi$ junction.  Here the Hamiltonian simplifies to 
\begin{eqnarray}
  H_{123} \propto -i\Gamma \cos\left(\frac{\phi_C-\phi_B}{2}\right)\gamma_3(\gamma_2-\gamma_1).
  \label{H123pijunction}
\end{eqnarray}
It follows that the linear combination $(\gamma_1+\gamma_2)/\sqrt{2}$ remains a zero-energy Majorana mode, while $\gamma_3$ and $(\gamma_1-\gamma_2)/\sqrt{2}$ combine into a finite-energy fermion.  While here the zero-energy Majorana carries weight only on the horizontal wires which formed the $\pi$ junction, in general its wavefunction will have weight on all three segments.  

As we braid Majorana fermions using the methods described in the main text, it is imperative that we avoid generating spurious zero-modes at the T-junction.  The above discussion implies that we are safe in this regard so long as we avoid $\pi$ junctions.  Fortunately, the semiconducting wires we considered naturally avoid such situations, since two wires at right angles to one another exhibit effective $p$-wave phases that differ by $\pi/2$ as discussed in the main text.  

\subsection{Wavefunction approach to Majorana fermion exchange}

In this section we explore in much greater detail the exchange of Majorana fermions in a 1D wire networks.  We once again emphasize the nontrivial nature of the problem since non-Abelian statistics in a 2D $p+ip$ superconductor is typically understood as arising because of superconducting vortices.  One might worry that the Majoranas in the wires perhaps bind vortices in the neighboring parent $s$-wave superconductor, but this is certainly not the case.  This becomes apparent when one recalls how the effective superconducting Hamiltonian for the wire is derived (see, \emph{e.g.}, Ref.\ \onlinecite{SauReview}).  Namely, one considers a Hamiltonian of the form $H = H_{\rm wire} + H_{\rm SC} + H_t$, where $H_{\rm wire}$ and $H_{\rm SC}$ describe the wire and superconductor in isolation, and $H_t$ encodes single-electron tunneling between the two.  Upon integrating out the gapped superconducting degrees of freedom assuming a \emph{uniform} pair field $\Delta$, one arrives at an effective Hamiltonian for the wire which includes proximity induced pairing terms.  Any phase variations in the parent superconductor's order parameter are ruled out by assumption, yet Majoranas can nevertheless exist in the wires.  Thus developing a physical picture for the exchange in this setting poses an extremely important issue.  

Our aim here is to provide greater rigor and a complementary picture for the discussion presented in the main text for how the Majoranas transform under exchange.  We will begin by constructing the many-body wavefunctions for a general 1D wire network supporting an arbitrary number of Majorana fermions.  We will then establish some important general results for braiding that rely on minimal assumptions about the underlying Hamiltonian.  Here there will be some overlap with the approach followed by Bonderson \emph{et al}.\cite{BondersonNonAbelianStatistics}, who recently revisited the issue of non-Abelian statistics in the fractional quantum Hall context.  Following this general analysis, we will consider again the exchange of Figs.\ \ref{Tjunction}(a)-(d) and explore how the Majoranas transform when the braid is implemented by keeping the superconducting phases fixed, as would be the case in practice.  Next, we will turn to an analysis of the exchange outlined in Figs.\ \ref{Tjunction}(e)-(h) and show that, as claimed in the main text, this braid transforms the Majoranas in an identical fashion to the braid of Figs.\ \ref{Tjunction}(a)-(d).  Finally, we will consider some special examples where the full many-body wavefunctions can be analyzed during the exchange, allowing us to explore important issues such as the overall phase acquired by the ground states under braiding.

\subsubsection{Construction of degenerate ground state wavefunctions}

Consider a 1D wire network with $2M$ well-separated, localized Majorana modes corresponding to operators $\gamma_{1,\ldots,2M}$ that satisfy $\gamma_j = \gamma_j^\dagger$ and $\gamma_j^2 = 1$.  By `well-separated', we mean that different Majorana wavefunctions overlap negligibly with each other.  Suppose moreover that the pairs $\gamma_{2j-1}$ and $\gamma_{2j}$ were germinated from the vacuum.  (For example, one could start with a non-topological network, generate $\gamma_1$ and $\gamma_2$ by nucleating a single topological region, then create $\gamma_3$ and $\gamma_4$ by forming another far-away topological region, \emph{etc.})  One can construct $M$ fermion operators from these via 
\begin{equation}
  f_j = \frac{1}{2}(\gamma_{2j-1} + i\gamma_{2j}),
  \label{fj}
\end{equation}
which correspond to zero-energy modes (up to corrections which are exponentially small in the separation between Majoranas).  These modes give rise to $2^M$ degenerate ground states which can be labeled by the occupation numbers $n_j = 0,1$ for the $f_j$ fermions.  We would like to construct these degenerate ground state wavefunctions and understand the exchange of Majorana fermions in 1D wire networks from this perspective.

Let us denote the positive-energy Bogoliubov-de Gennes quasiparticle operators by $d_{\alpha}$, each of which must annihilate the ground states.  As usual, the explicit form of these ground states is nontrivial because both the Majorana operators and $d_\alpha$ represent linear combinations of the original fermion creation and annihilation operators for the 1D wire network.  By construction the wavefunction 
\begin{equation}
  |\psi\rangle = \frac{1}{\sqrt{\mathcal{N}}}\prod_\alpha d_\alpha |{\rm vac}\rangle, 
\end{equation}
with $|{\rm vac}\rangle$ the vacuum of the original fermion operators and $\mathcal{N}$ the normalization, must constitute one of the degenerate ground states since any $d_\alpha$ clearly annihilates this state.  Because we pulled $\gamma_{2j-1}$ and $\gamma_{2j}$ out of the vacuum, we are guaranteed that $|\psi\rangle$ will be an eigenvector of $f_j^\dagger f_j$ with eigenvalue $n_j$.  

How we label this state in terms of the $n_j$ occupation numbers is a matter of convention because there is a sign ambiguity in the Majorana wavefunctions.  Specifically, if $\phi_j$ is the Majorana wavefunction corresponding to $\gamma_j$, then $-\phi_j$ also denotes a legitimate Majorana wavefunction that preserves the relations $\gamma_j = \gamma_j^\dagger$ and $\gamma_j^2 = 1$.  As an example, suppose that upon making some particular overall sign choices for $\phi_{1,2}$ we find that $f_1|\psi\rangle = 0$; we would then identify $|\psi\rangle$ with $n_1 = 0$.  Had we chosen the opposite sign for $\phi_2$, however, we would find instead that $f_1^\dagger|\psi\rangle = 0$ and identify this state with $n_1 = 1$ (sending $\phi_2 \rightarrow -\phi_2$ is equivalent to sending $\gamma_2 \rightarrow -\gamma_2$ and $f_1 \rightarrow f_1^\dagger$).  We will assume for concreteness that the signs of the Majorana wavefunctions have been chosen such that $|\psi\rangle$ corresponds to the ground state with $n_j = 1$, \emph{i.e.}, 
\begin{equation}
  |11\cdots 1\rangle = |\psi\rangle.
\end{equation}
The ground state with $n_j = 0$ can then be written
\begin{equation}
  |00\cdots 0\rangle = f_1 f_2\cdots f_M |\psi\rangle,
\end{equation}
which is manifestly annihilated by all $f_j$.  All other ground states can be obtained by applying creation operators $f_j^\dagger$ to $|00\cdots 0\rangle$ or, equivalently, annihilation operators $f_j$ to the state $|11\cdots 1\rangle$.  
 
Before moving on, we note that the way we combined Majoranas in Eq.\ (\ref{fj}) to construct the ordinary $f_j$ fermion operators was convenient because of how we assumed the Majoranas were germinated, but is by no means unique.  One could always choose to combine pairs of Majoranas differently and construct operators $f_j'$ and ground states $|n_1' n_2'\cdots n_M'\rangle$.  The ground states in this representation would then be related to the ground states we defined above simply by a change of basis.  

\subsubsection{General results for Majorana exchange}
\label{GeneralResults}

We now proceed to obtain some important generic results for Majorana exchange that rely on very minimal assumptions about the underlying Hamiltonian for the 1D wire network.  Suppose that we wish to exchange $\gamma_1$ and $\gamma_2$.  [Below it will prove extremely convenient to work in a basis where the Majoranas we braid combine into an ordinary fermion.  This is the case for the basis we introduced above, since $f_1 = (\gamma_1+i\gamma_2)/2$.  If instead we wanted to exchange, say, $\gamma_1$ and $\gamma_3$, one would first want to change basis and write the ground states in terms of operators such as $f_1' = (\gamma_1 + i \gamma_3)/2$ and $f_2' = (\gamma_2 + i \gamma_4)/2$, then proceed as we outline below.]  Let $\lambda$ be the parameter in the Hamiltonian that varies to implement the exchange of $\gamma_1$ and $\gamma_2$.  If $\lambda$ varies from $\lambda_{{\rm init}}$ to $\lambda_{\rm final}$ during the course of the exchange, then we require that $H(\lambda_{\rm final}) = H(\lambda_{\rm init})$ so that the Hamiltonian returns to its original form after the braid.  (If the Hamiltonian does not return to its original form, then we can not make rigorous statements about how the wavefunctions transform under the exchange.)

There are two important contributions that one must understand to analyze the exchange.  The first are the Berry phases acquired by the degenerate ground states, which follow from
\begin{eqnarray}
  \langle n_1 n_2\cdots n_M|\partial_\lambda| m_1 m_2\cdots m_M\rangle.
\end{eqnarray}
We have included the possibility that non-trivial off-diagonal Berry phases occur, since \emph{a priori} these need not vanish.  If we assume that the degenerate ground states return to their precise original form at $\lambda_f$, the Berry phases encode all information about the exchange.  However, if we relax this assumption (which will indeed be useful below), then we additionally need to compare the explicit changes between the initial and final states.  In general, the final ground states could represent a nontrivial linear combination of the initial ground states, so we write
\begin{eqnarray}
  |n_1 n_2 \cdots n_M\rangle_f &=& \sum_{\{m_j\}} \rho(\{m_j\};\{n_j\})|m_1 m_2 \cdots m_M\rangle_i,
  \nonumber \\
\end{eqnarray}
where the subscripts $i/f$ denote the initial/final states.  The combination of the Berry phases and the coefficients $\rho(\{m_j\};\{n_j\})$ fully specify the outcome of the exchange.  

While the problem appears daunting, a remarkable amount of progress can in fact be made on very general grounds.  We will make only one additional assumption about the physical system.  Specifically, to carry out the exchange we will assume that only local terms in the Hamiltonian---such as local chemical potentials---need to be modified, and that such modifications only impact the Majoranas $\gamma_{1,2}$ which are being braided.  This will certainly be the case, for example, in the eight-Majorana configuration from Fig.\ 4(a) of the main text; in fact, there any pair can be exchanged without disturbing any other Majoranas in the system.  However, one is not guaranteed that this is always immediately possible, since there may be intervening Majoranas that prevent $\gamma_1$ and $\gamma_2$ from being so exchanged.  Such cases can be treated in one of two ways.  First, one can always first transport the Majoranas in the system (but importantly \emph{without} braiding any of them) to produce an arrangement where $\gamma_{1,2}$ can be directly exchanged.  Alternatively, one can always pairwise braid Majoranas in the system until the intervening Majoranas are removed.  We will proceed below by assuming that in such cases, the prerequisite translations or exchanges have already been carried out.  

We will first demonstrate that all off-diagonal Berry phase elements vanish trivially in the basis we have chosen, and that the exchange depends only on the occupation number $n_1$ for the $f_1 = (\gamma_1+i\gamma_2)/2$ fermion.  This follows from our assumption above that the exchange can be implemented by simply adjusting local terms in the Hamiltonian that affect only the regions near $\gamma_{1,2}$.  Assuming all other Majoranas are well-localized far from $\gamma_{1,2}$, their wavefunctions will be unaffected by these local changes to the Hamiltonian (up to exponentially small corrections).  Consequently, to an excellent approximation $f_{2,\ldots,M}$ can be assumed $\lambda$-independent.  

In contrast, $f_1$ and the bulk quasiparticle operators $d_\alpha$ do depend on $\lambda$.  Explicitly displaying the $\lambda$-dependence, the ground state $|00\cdots0\rangle$ can then be written as 
\begin{eqnarray}
  |00\cdots0\rangle = f_1(\lambda)f_2\cdots f_M|\psi(\lambda)\rangle.
  \label{0000}
\end{eqnarray}
Again, all other ground states can be obtained from this by application of $f_1(\lambda)^\dagger$ and $f_{2,\ldots,M}^\dagger$.  Because $f_{2,\ldots,M}$ are $\lambda$-independent, differentiating the states $|n_1 n_2\cdots n_M\rangle$ with respect to $\lambda$ does not change the occupation numbers for the corresponding zero-energy modes.  It therefore follows that 
\begin{eqnarray}
  \langle n_1 n_2 \cdots n_M|\partial_\lambda|m_1 m_2 \cdots m_M\rangle \propto \delta_{n_2,m_2}\cdots \delta_{n_M,m_M}.
\end{eqnarray}
One can see this relation formally by inserting the ground states obtained from Eq.\ (\ref{0000}) into the left-hand side of the above equation.  We furthermore have
\begin{eqnarray}
  \langle n_1 n_2 \cdots n_M|\partial_\lambda|m_1 n_2 \cdots n_M\rangle \propto \delta_{n_1,m_1}
\end{eqnarray}
since with $n_1\neq m_1$ the above bra and ket are trivially orthogonal because they exhibit different fermion parity.  Thus all off-diagonal Berry phases indeed vanish, and the only possibly non-zero Berry phases follow from
\begin{eqnarray}
  \Lambda_0(\lambda) &\equiv& \langle 0 n_2 \cdots n_M|\partial_\lambda |0 n_2 \cdots n_M \rangle 
  \nonumber \\
  &=& \langle \psi(\lambda)|f_1(\lambda)^\dagger\partial_\lambda f_1(\lambda)|\psi(\lambda)\rangle,
  \label{Lambda0}
  \\
  \Lambda_1(\lambda) &\equiv& \langle 1 n_2 \cdots n_M|\partial_\lambda |1 n_2 \cdots n_M \rangle 
  \nonumber \\
  &=& \langle \psi(\lambda)|\partial_\lambda |\psi(\lambda)\rangle,
  \label{Lambda1}
\end{eqnarray}
which manifestly depend only on $n_1$.  As one would intuitively expect, the additional far-away Majorana modes do not impact the outcome of the exchange of $\gamma_{1,2}$.  These modes trivially factor out in the above sense, and the exchange simply sends $\gamma_i \rightarrow \gamma_i$ for $i \neq$ 1 or 2.  

We can actually evaluate the Berry phases even further.  From Eqs.\ (\ref{Lambda0}) and (\ref{Lambda1}), we obtain the following relation:
\begin{eqnarray}
  \Lambda_0(\lambda) = \Lambda_1(\lambda) + \langle \psi(\lambda)|f_1(\lambda)^\dagger [\partial_\lambda f_1(\lambda)]|\psi(\lambda)\rangle,
  \label{RelativeBerryPhase}
\end{eqnarray}
where on the right side the derivative acts only on $f_1(\lambda)$.  One can always decompose $\partial_\lambda f_1(\lambda)$ as
\begin{eqnarray}
  \partial_\lambda f_1(\lambda) &=& A(\lambda) f_1(\lambda) + A'(\lambda) f_1(\lambda)^\dagger 
  \nonumber \\
  &+& \sum_\alpha\left[B_\alpha(\lambda)d_\alpha(\lambda) + B_\alpha'(\lambda) d_\alpha(\lambda)^\dagger\right].
  \label{dlambdaf}
\end{eqnarray}
(The right-hand side does not involve $f_{2,\ldots,M}$ because by assumption the corresponding wavefunctions have negligible weight near $\gamma_1$ and $\gamma_2$.)  Inserting this expression into Eq.\ (\ref{RelativeBerryPhase}), one finds that only the $A(\lambda)$ term contributes:
\begin{eqnarray}
  \Lambda_0(\lambda) = \Lambda_1(\lambda) + A(\lambda) .
  \label{RelativeBerryPhase2}
\end{eqnarray}
It is useful to isolate $A(\lambda)$ in Eq.\ (\ref{dlambdaf}) by considering the following anticommutator,
\begin{equation}
  \{f_1(\lambda)^\dagger,\partial_\lambda f_1(\lambda)\} = A(\lambda).
  \label{A}
\end{equation}
Suppose now that throughout the exchange the following standard relations hold
\begin{eqnarray}
  f_1(\lambda) &=& \frac{1}{2}[\gamma_1(\lambda)+i\gamma_2(\lambda)]
  \nonumber \\
  \gamma_j(\lambda) &=& \gamma_j(\lambda)^\dagger
  \label{MajoranaRelations} \\
  \gamma_j(\lambda)^2 &=& 1.
  \nonumber
\end{eqnarray}
Note that this does not constitute an additional physical assumption, but rather a convention; one can always choose the operators in this way.  The anticommutator in Eq.\ (\ref{A}) is then
\begin{eqnarray}
  \{f_1(\lambda)^\dagger,\partial_\lambda f_1(\lambda) \} &=& 
  \frac{1}{4}\big{[}\{\gamma_1,\partial_\lambda \gamma_1\} + \{\gamma_2,\partial_\lambda \gamma_2\} 
  \nonumber \\
  &+&i\{\gamma_1,\partial_\lambda \gamma_2\} - i \{\gamma_2,\partial_\lambda \gamma_1\}\big{]},
  \label{anticommutator}
\end{eqnarray}
where for brevity on the right-hand side we suppressed the $\lambda$ dependence.
The first two terms vanish trivially because
\begin{eqnarray}
  \{\gamma_j,\partial_\lambda \gamma_j\} &=& \gamma_j(\partial_\lambda\gamma_j) + (\partial_\lambda \gamma_j)\gamma_j 
  \nonumber \\
  &=& \partial_\lambda(\gamma_j\gamma_j) = \partial_\lambda(1) = 0.
\end{eqnarray}
Interestingly, the last two terms in Eq.\ (\ref{anticommutator}) \emph{also} vanish (up to exponentially small corrections) because $\gamma_1$ and $\gamma_2$ are localized far apart from one another.  (To see this explicitly, one can expand $\gamma_{1,2}$ in terms of the original fermion creation and annihilation operators for the wire network, evaluate $\partial_\lambda \gamma_{1,2}$ in terms of this expansion, and then evaluate the anticommutators.)  Thus we obtain the remarkable result that as long as Eqs.\ (\ref{MajoranaRelations}) hold throughout the exchange, 
\begin{equation}
  \Lambda_1(\lambda) = \Lambda_0(\lambda)
\end{equation}
and the states with $n_1 = 0$ and 1 acquire identical Berry phases under the exchange.  

This does not imply that the exchange is trivial, since nontrivial effects can still arise due to explicit differences between the initial and final states.  Recall that by assumption the Hamiltonian returns back to its original form after the exchange.  Therefore we must have $f_1(\lambda_{\rm final}) = e^{i\theta_f}f_1(\lambda_{\rm init})$ and $d_\alpha(\lambda_{\rm final}) = e^{i\theta_\alpha}d_\alpha(\lambda_{\rm init})$, for some unspecified phase factors $\theta_f$ and $\theta_\alpha$.  By choosing the bulk quasiparticle wavefunctions appropriately, we can always set without loss of generality $\theta_\alpha = 0$, though we must allow for nontrivial $\theta_f$ since we restricted $f_1(\lambda)$ to satisfy Eqs.\ (\ref{MajoranaRelations}) for all $\lambda$.  The initial and final ground states are then related by
\begin{eqnarray}
  |0n_2 \cdots n_M\rangle_f &=& e^{i\theta_f}|0n_2 \cdots n_M\rangle_i
  \nonumber \\
  |1 n_2 \cdots n_M\rangle_f &=& |1 n_2 \cdots n_M\rangle_i.
  \label{ExplicitPhaseDifference}
\end{eqnarray}

The most important implication of these results is that the outcome of the exchange follows solely from the evolution of the localized Majorana wavefunctions for $\gamma_{1,2}$, which determines $\theta_f$.  [This is true up to an overall Berry phase arising from $\Lambda_j(\lambda)$ which, by contrast, depends additionally on the evolution of the bulk quasiparticle states.  But this contribution is non-universal in any event as we will see later.]  It is worth emphasizing that the wavefunctions corresponding to $\gamma_{1,2}$ depend neither on the presence of the additional Majoranas in the system, nor on changes to the bulk quasiparticle wavefunctions far from $\gamma_{1,2}$.  This leads to a great simplification: if we know how $\gamma_{1,2}$ transform in some minimal configuration, then the same transformation must hold when arbitrarily many additional distant Majoranas are introduced by, say, nucleating other topological regions into the network.  Once this is known we can determine the unitary operator $U_{12}$ that acts on $\gamma_{1,2}$ to implement their exchange.  Importantly, $U_{12}$ is basis independent as it only acts on the Majorana operators, and therefore allows one to deduce the evolution of the many-body ground states under the exchange (up to an overall phase) in any basis one chooses.  

In the more general situation where one exchanges $\gamma_i$ and $\gamma_j$, exactly the same analysis holds when one works in a basis where one of the occupation numbers corresponds to the ordinary fermion operator $f' = (\gamma_i + i\gamma_j)/2$, as briefly mentioned at the beginning of this subsection.  The implications discussed in the preceding paragraph that follow from these results also carry over to this more general case.  We stress that although it is easiest to draw these conclusions by choosing a different basis depending on which Majoranas are being braided, one again ultimately deduces basis-independent operators $U_{ij}$ that act on $\gamma_{i,j}$ to implement their braid.  The set of unitary operators $\{U_{ij}\}$ are sufficient to deduce how the degenerate ground states transform in a fixed basis under arbitrary exchanges (up to an overall non-universal phase); see for example Eq.\ (9) in Ref.\ \onlinecite{Ivanov}.  

We have now distilled the problem of demonstrating non-Abelian statistics down to that of determining how the Majoranas $\gamma_1$ and $\gamma_2$ transform under the exchanges shown in Fig.\ \ref{Tjunction}.  These two elementary operations alone indeed allow one to braid arbitrary pairs of Majorana fermions in networks composed of trijunctions, such as that shown in Fig.\ 4(a) of the main text.  We proceed now by analyzing these two cases in turn.

\subsubsection{Exchange of $\gamma_{1,2}$ from Figs.\ \ref{Tjunction}(a)-(d)}

\label{Fig1exchangeA}

\begin{figure}
\centering{
  \includegraphics[width=3in]{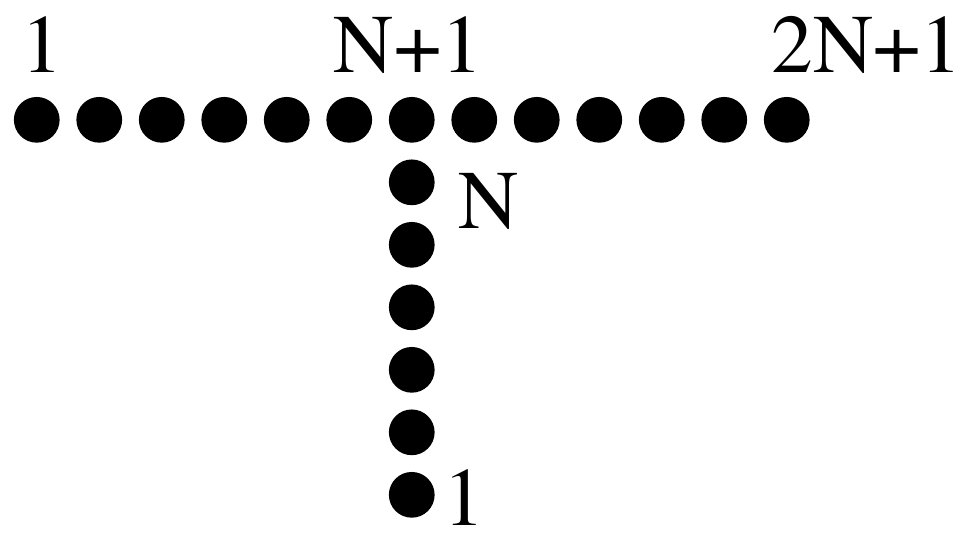}
  \caption{Lattice structure giving rise to the T-junction.}
  \label{LatticeTjunction}}
\end{figure}

Consider the T-junction in Fig.\ \ref{Tjunction}(a).  As in the main text we will describe each wire by Kitaev's toy lattice model.  (We remind the reader that the experimentally realistic semiconductor wire Hamiltonian for the junction can be smoothly deformed to this minimal model without closing the gap.  We choose to work with the latter since it is more convenient, and are free to do so because exchange statistics is a universal property that is insensitive to details of the Hamiltonian as long as the gap remains finite.)  We take the horizontal and vertical wires to consist of $2N+1$ and $N$ sites as Fig.\ \ref{LatticeTjunction} illustrates, and denote the spinless fermion operators for the horizontal and vertical chains respectively by $c_x$ and $\tilde c_y$.  For simplicity we will take the pairing and nearest-neighbor hopping amplitudes to be equal, which allows us to conveniently express the Hamiltonian as
\begin{eqnarray}
  H &=& -\sum_{x = 1}^{2N+1}\mu_x c_x^\dagger c_x - \sum_{y = 1}^N \tilde \mu_y \tilde c_y^\dagger \tilde c_y
  \nonumber \\
  &+&\sum_{x = 1}^{2N}t_x(e^{-i\phi/2}c_x^\dagger +e^{i\phi/2}c_x)(e^{-i\phi/2}c_{x+1}^\dagger-e^{i\phi/2}c_{x+1})
  \nonumber \\
  &+&\sum_{y = 1}^{N}\tilde t_y(e^{-i\tilde \phi/2}\tilde c_y^\dagger +e^{i\tilde \phi/2}\tilde c_y)(e^{-i\tilde \phi/2}\tilde c_{y+1}^\dagger-e^{i\tilde \phi/2}\tilde c_{y+1})
  \nonumber \\
  \label{Hjunction}
\end{eqnarray}
where we identify $\tilde c_{N+1} \equiv c_{N+1}$.  

It is important to recall from the discussion below Eq.\ (\ref{Trijunction}) that the superconducting phases for the T-junction are only well-defined with respect to a direction in each wire.  We have chosen to write the pairing terms above as $-t_x e^{i\phi}c_x c_{x+1} + h.c.$ and similarly for the $\tilde\phi$ terms.  The former of course equals $-t_x e^{i(\phi+\pi)}c_{x+1} c_{x} + h.c.$, so one can always equivalently refer to the superconducting phase in the horizontal wire as $\phi$ with respect to the `right' direction or $\phi + \pi$ with respect to the `left' direction.  Similarly, one can label the superconducting phase for the vertical wire as either $\tilde\phi$ with respect to the `up' direction or $\tilde\phi+\pi$ with respect to the `down' direction.  One can even divide up each wire into distinct regions and use a different convention in each one.  We will indeed find this useful to do here.  Throughout this subsection we will employ a convention where the superconducting phases are labeled as $\phi+\pi$ in the left half of the horizontal wire, $\phi$ in the right half, and $\tilde\phi+\pi$ in the vertical wire.  This will prove extremely convenient for describing the Majorana operators as we will see shortly.  Note also that here and in the following subsection we will assume that $\phi$ and $\tilde\phi$ do not differ by an integer multiple of $\pi$ so that we avoid generating $\pi$ junctions during the exchange.  

In the initial configuration the vertical wire is non-topological, while the horizontal wire is topological and thus exhibits end-Majoranas $\gamma_{1,2}$ which we would like to exchange counterclockwise as in Fig.\ \ref{Tjunction}(a)-(d).  Given our discussion in the previous subsection, we will choose $\gamma_{1,2}$ to satisfy the usual relations $\gamma_{j} = \gamma_j^\dagger$ and $\gamma_j^2 = 1$ throughout, for in this case we need only compare the initial and final Majorana operators to deduce the outcome of the exchange (up to an overall non-universal phase).  In the main text we already deduced how the Majoranas transform under this exchange, but here we will rederive this transformation rule in the case where the superconducting phases are held fixed.  In other words, we now wish to braid $\gamma_{1,2}$ by only varying $\mu_x, \tilde\mu_y, t_x$, and $\tilde t_y$.  To implement the exchange we take these couplings to depend on a parameter $\lambda$ that varies from $\lambda_a$, corresponding to the initial setup of Fig.\ \ref{Tjunction}(a), to $\lambda_d$, corresponding to the final configuration of Fig.\ \ref{Tjunction}(d).  The intermediate values $\lambda_{b,c}$ will correspond to Figs.\ \ref{Tjunction}(b) and (c).  

We will model the initial setup at $\lambda = \lambda_a$ by assuming for simplicity that  $t_x = t>0$, $\mu_x = \tilde t_y = 0$, and $\tilde \mu_y = \mu<0$.  We can then read off the form of the initial Majorana operators from Eq.\ (\ref{Hjunction}):
\begin{eqnarray}
  \gamma_1(\lambda_a) &=& e^{-i(\phi+\pi)/2}c_1^\dagger + h.c.
  \label{gamma1a}
  \\
  \gamma_2(\lambda_a) &=& e^{-i\phi/2}c_{2N+1}^\dagger + h.c.
  \label{gamma2a}
\end{eqnarray}
since these combinations are explicitly absent from the initial Hamiltonian.  Note that the overall signs of $\gamma_{1,2}(\lambda_a)$ have been chosen completely arbitrarily above.  Suppose now that we adiabatically transport $\gamma_1$ rightward and downward, leading to the configuration of Fig.\ \ref{Tjunction}(b).  For example, $\gamma_1$ can be transported rightward one site by taking $\mu_1 = \lambda \mu$ and $t_1 = (1-\lambda)t$ and varying $\lambda$ from 0 to 1.  Upon similarly transporting $\gamma_1$ all the way to site 1 of the vertical chain, at $\lambda = \lambda_b$ we end up with $\mu_x = \mu$, $t_x = 0$ in the non-topological region of the horizontal wire, $\mu_x = 0$ and $t_x = t>0$ in the topological region of the horizontal wire, and $\tilde \mu_y = 0$ and $\tilde t_y = t$.  Once again, we can then read off the Majorana operators,
\begin{eqnarray}
  \gamma_1(\lambda_b) &=& s_1\left[e^{-i(\tilde\phi+\pi)/2}\tilde c_1^\dagger + h.c.\right]
  \\
  \gamma_2(\lambda_b) &=& \gamma_2(\lambda_a),
  \label{gamma2b}
\end{eqnarray}
where $s_1 = \pm 1$.  At this point the utility of our convention for the superconducting phases becomes clear: with this labeling scheme, the phase factors appearing in Eqs.\ (\ref{gamma1a}) to (\ref{gamma2b}) follow directly from the superconducting phase `felt' locally by the Majoranas.  

In contrast to the overall signs of $\gamma_{1,2}(\lambda_a)$ above, $s_1$ and the sign of $\gamma_2(\lambda_b)$ are \emph{not} arbitrary since these operators evolve adiabatically from $\gamma_{1,2}(\lambda_a)$.  Since we did not transport $\gamma_2$ in this step, clearly this operator must remain invariant as we have written.  Determining the sign $s_1$ is, however, trickier.  One can in principle determine $s_1$ by tracking the evolution of the Majorana wavefunction by explicit calculation.  This brute force approach does not provide much insight, however, so we will deduce the sign $s_1$ by different means.  Specifically, we will ask what happens when we go from Fig.\ \ref{Tjunction}(a) to (b) in a slightly deformed Hamiltonian which smoothly connects to the one we discussed above.  

Suppose that instead of the superconducting phases jumping discontinuously between the horizontal and vertical wires, the phase in the latter varied spatially from $\tilde\phi+\pi$ at the bottom to $\phi+\pi$ at the top.  (All the phase variation can happen very locally near the trijunction, so long as it is smooth.)  Let us denote the Majoranas in this deformed problem by $\overline\gamma_{1,2}(\lambda)$, which can be expressed as
\begin{eqnarray}
  \overline\gamma_1(\lambda) &=& e^{-i\phi_1(\lambda)/2}C_1(\lambda)^\dagger + h.c.
  \label{phi1}
  \\
  \overline\gamma_2(\lambda) &=& e^{-i\phi_2(\lambda)/2}C_2(\lambda)^\dagger + h.c.,
  \label{phi2}
\end{eqnarray}
where $C_{1,2}(\lambda)$ correspond to the fermion operators for the sites where $\overline\gamma_{1,2}$ are localized at the `time' $\lambda$ of the exchange.  The phases $\phi_{1,2}(\lambda)$ are simply given by the superconducting phases felt locally by the Majoranas, and thus rotate as we transport $\overline\gamma_{1,2}$.  Upon going from Fig.\ \ref{Tjunction}(a) to (b), $\phi_2(\lambda)$ of course remains unchanged while $\phi_1(\lambda)$ rotates from $\phi + \pi$ to $\tilde\phi + \pi$.  Whether this rotation happens clockwise or counterclockwise, however, makes all the difference: if in the former case $e^{-i(\phi+\pi)/2}\rightarrow e^{-i(\tilde\phi + \pi)/2}$, then in the latter $e^{-i(\phi+\pi)/2}\rightarrow -e^{-i(\tilde\phi + \pi)/2}$.  This follows directly from the fact that $e^{-i\phi_j(\lambda)/2}$ changes sign upon shifting $\phi_j$ by $2\pi$.  Because of this mathematical property, it is useful to introduce branch cuts in the space of $\phi_j(\lambda)$, precisely as in Ivanov's construction \cite{Ivanov}, so that the Majorana operators are single-valued away from the cut but change sign when $\phi_j(\lambda)$ crosses the cut.  Arbitrarily, we will take the branch cut to occur at $\phi_j = 0$.  

\begin{figure}
\centering{
  \includegraphics[width=2.5in]{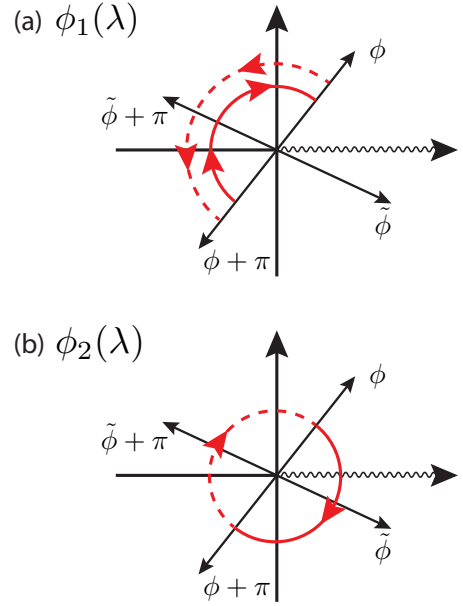}
  \caption{Trajectory of (a) $\phi_1(\lambda)$ and (b) $\phi_2(\lambda)$ defined in Eqs.\ (\ref{phi1}) and (\ref{phi2}), which are the superconducting phases `felt' locally by the braided Majoranas $\overline\gamma_1$ and $\overline\gamma_2$ of the deformed Hamiltonian described in the text.  The solid red lines correspond to the trajectories obtained for the type of exchange illustrated in Figs.\ \ref{Tjunction}(a)-(d), while the dashed red lines correspond to Figs.\ \ref{Tjunction}(e)-(h).  Because the Majorana operators depend on $e^{i\phi_j(\lambda)/2}$, we introduce a branch cut along $\phi_j = 0$ so that $\overline\gamma_{1,2}$ are single-valued away from the cut but change sign when $\phi_j$ crosses the branch cut.  In the exchange from Figs.\ \ref{Tjunction}(a)-(d), both $\phi_1$ and $\phi_2$ always rotate by $\pi$ with the same orientation, and hence only one of the two Majorana operators acquires a minus sign due to the branch cut.  In the exchange of Figs.\ \ref{Tjunction}(e)-(h), however, $\phi_1$ and $\phi_2$ rotate by $\pi$ with \emph{opposite} orientations, so either both $\overline\gamma_{1,2}$ acquire a minus sign due to the branch cut or neither do.  There is an additional minus sign though that $\overline\gamma_2$ picks upon crossing the trijunction between Figs.\ \ref{Tjunction}(f) and (g), so the net result is that in both exchanges one of the two Majoranas changes sign, just like for vortices in a 2D $p+ip$ superconductor.  Since the deformed model and the Hamiltonian of interest can be smoothly connected without a gap closure, the same is true for the original Majoranas, $\gamma_{1,2}$.   }
  \label{BranchCuts}}
\end{figure}

If one is to avoid generating spurious zero modes at a $\pi$ junction during this step (and thus smoothly connect to the problem of interest), there is a unique orientation in which the superconducting phase in the vertical wire can vary.  Namely, it must vary from $\tilde\phi+\pi$ at the bottom to $\phi+\pi$ at the top without passing through $\phi$.  [To understand why going through $\phi$ would generate a $\pi$ junction with zero modes, recall the discussion below Eq.\ (\ref{Trijunction}), keeping in mind our labeling conventions.]  Consequently, $\phi_1(\lambda)$ rotates from $\phi+\pi$ at $\lambda = \lambda_a$ to $\tilde\phi + \pi$ at $\lambda_b$, also without passing through $\phi$.  In Fig.\ \ref{BranchCuts}(a) for example, the rotation happens clockwise as shown by the solid red line.  Since the deformed problem smoothly connects to the physical problem of interest throughout this step of the exchange, we can now easily deduce the sign $s_1$ that determines $\gamma_1(\lambda_b)$: $s_1 = -1$ if $\phi_1(\lambda)$ crosses the branch cut while $s_1 = +1$ otherwise.  

Next, let us return to the original problem and similarly transport $\gamma_2$ leftward, producing the configuration of Fig.\ \ref{Tjunction}(c).  At $\lambda = \lambda_c$, we again have $\mu_x = \mu<0$, $t_x = 0$ in the non-topological region of the horizontal wire, $\mu_x = 0$ and $t_x = t>0$ in the topological region of the horizontal wire, and $\tilde \mu_y = 0$ and $\tilde t_y = t$.  The Majorana operators then evolve to
\begin{eqnarray}
  \gamma_1(\lambda_c) &=& \gamma_1(\lambda_b)
  \\
  \gamma_2(\lambda_c) &=& s_2\left[e^{-i(\phi+\pi)/2}c_1^\dagger + h.c.\right]
  \label{gamma2c}
\end{eqnarray}
for some sign $s_2$.  

To determine $s_2$, we again turn to a deformed problem which provides useful insight and eschews the need for brute force calculation.  As before, we seek a companion Hamiltonian with Majoranas $\overline\gamma_{1,2}$ defined as in Eq.\ (\ref{phi1}) and (\ref{phi2}), where the superconducting phase $\phi_2(\lambda)$ felt locally by $\overline\gamma_2$ varies smoothly as $\overline\gamma_2$ shuttles leftward.  This can be achieved by deforming the original Hamiltonian so that the superconducting phase in the horizontal wire varies spatially from $\phi+\pi$ on the left, to $\tilde\phi$ in the center where it meets the vertical wire, and then to $\phi$ on the right.  There is again a unique orientation in which this variation can take place if this problem is to smoothly connect with the original one---the superconducting phase must never pass through $\tilde\phi+\pi$ to avoid generating a $\pi$ junction in the topological region during this step.  Given how the horizontal wire's superconducting phase varies in the deformed Hamiltonian, it follows that $\phi_2(\lambda)$ must rotate from $\phi$ at $\lambda_b$ to $\phi+\pi$ at $\lambda_c$ in an orientation that passes through $\tilde\phi$.  In the example of Fig.\ \ref{BranchCuts}(b), this implies that $\phi_2$ rotates clockwise as shown by the solid red line, crossing the branch cut (wavy line) in the process.  We can now trivially extract the sign $s_2$ using the fact that the deformed Hamiltonian and the physical Hamiltonian of interest connect smoothly to one another throughout this step: $s_2 = -1$ if $\phi_2(\lambda)$ crosses the branch cut but $s_2 = +1$ otherwise.  

Going back to the original problem, let us complete the exchange by transporting $\gamma_1$ up and to the right, producing the configuration of Fig.\ \ref{Tjunction}(d).  The Hamiltonian returns to its original form specified above, and the Majoranas evolve to
\begin{eqnarray}
  \gamma_1(\lambda_d) &=& s_1's_1\left[e^{-i\phi/2}c_{2N+1}^\dagger + h.c.\right]
  \\
  \gamma_2(\lambda_d) &=& \gamma_2(\lambda_d)
\end{eqnarray}
for some sign $s_1'$.  

One can deduce $s_1'$ using the same approach employed to find $s_1$.  We analogously introduce a deformed Hamiltonian where the superconducting phase varies spatially in the vertical wire from $\tilde\phi+\pi$ at the bottom to $\phi$ at the top, without passing through $\phi+\pi$ to avoid generating zero modes at a $\pi$ junction.  The Majorana $\overline\gamma_1$ then feels a local superconducting phase $\phi_1(\lambda)$ which rotates from $\phi+\pi$ at $\lambda_c$ to $\phi$ at $\lambda_d$ with an orientation that avoids passing through $\phi+\pi$.  In the example of Fig.\ \ref{BranchCuts}(a), this rotation happens clockwise as indicated by the solid red line.  Using continuity of the deformed problem and the physical problem we immediately deduce that $s_1' = -1$ if $\phi_1(\lambda)$ crosses the branch cut during this step and $s_1' = +1$ otherwise.  

At the end of the exchange we find that $\gamma_1(\lambda_d) = s_1' s_1 \gamma_2(\lambda_a)$ and $\gamma_2(\lambda_d) = s_2 \gamma_1(\lambda_a)$.  In Fig.\ \ref{BranchCuts} only the phase $\phi_2$ crosses the branch cut as one can see from the solid red lines; in this case $s_1 = s_1' = 1$ and $s_2 = -1$ so that one obtains $\gamma_1 \rightarrow -\gamma_2$ and $\gamma_2\rightarrow \gamma_1$.  More generally, it follows from our discussion above that $\phi_1$ and $\phi_2$ both always rotate by $\pi$ with the same orientation, and thus necessarily only one of the two crosses the branch cut.  Thus we obtain the result that the exchange sends
\begin{eqnarray}  
  \gamma_1 &\rightarrow& s \gamma_2
  \\
  \gamma_2 &\rightarrow& -s\gamma_1
\end{eqnarray}
with some sign $s$ that depends on $\phi, \tilde\phi$, and where one chooses the branch cut.  (We can always get rid of $s$ by absorbing this sign into the definition of $\gamma_2$.)  Thus we have derived the transformation rule for the exchange of Fig.\ \ref{Tjunction}(a)-(d) in a complementary manner to that of the main text.  

It is worth briefly contrasting these two approaches.  In this subsection we studied the exchange of two Majorana fermions when the superconducting phases in the wires were kept fixed.  We deduced how the Majoranas transformed here by considering a companion problem where the superconducting phases were allowed to change, but only \emph{locally} in the vicinity of the trijunction.  This approach required more extensive formalism but had the virtue of adhering closer to the physical situation for experiment.  It is also interesting that the picture that emerged is remarkably similar to Ivanov's picture for exchange of vortices in a 2D $p+ip$ superconductor\cite{Ivanov}, despite the absence of vortices in the wire network.  In the main text we instead deformed the Hamiltonian such that the superconducting phases in each segment of the network varied \emph{globally} during the exchange so as to keep the Hamiltonian real.  At the end, however, the superconducting order parameter changed sign, so to return the Hamiltonian back to its original form we reversed the sign of the pairing by sending $f^\dagger = (\gamma_1 - i\gamma_2)/2 \rightarrow i f^\dagger$.  This method had the virtue that all Berry phases explicitly vanished while the Majoranas were transported, allowing one to get to the answer very directly and with minimal formalism.  We stress that these two approaches are not unrelated, and it is interesting to note the connection between the deformed companion problem introduced in this section and the deformed Hamiltonian considered in the main text.  In the former case the superconducting phases felt by the Majoranas $\overline\gamma_{1,2}$ varied smoothly as they moved across the network, and as a result the phases $\phi_{1,2}$ smoothly rotated by $\pi$ during the course of the braid.  The latter case effectively corresponds to an extreme version of this wherein $\phi_{1,2}$ acquire the \emph{entire} $\pi$ rotation only at the very end of the exchange.

\subsubsection{Exchange of $\gamma_{1,2}$ from Figs.\ \ref{Tjunction}(e)-(h)}

We will now analyze the counterclockwise exchange of Figs.\ \ref{Tjunction}(e)-(h), where $\gamma_1$ and $\gamma_2$ reside on different topological regions of the network.  The same general approach applied in the previous subsection will be applied here.  As before, the T-junction will be described by $H$ in Eq.\ (\ref{Hjunction}), the Majoranas $\gamma_{1,2}$  will be defined so that $\gamma_{j} = \gamma_j^\dagger$ and $\gamma_j^2 = 1$ throughout, and their exchange will be implemented by taking the couplings $\mu_x, \tilde\mu_y, t_x$, and $\tilde t_y$ dependent on a parameter $\lambda$.  We will hold $\phi$ and $\tilde\phi$ fixed, but will again deduce the evolution of the Majoranas by considering a deformed companion Hamiltonian where the superconducting phases vary locally near the trijunction.  It will be convenient to employ a slightly different labeling convention for the superconducting phases here compared to the previous subsection: we will label these by $\phi$ in the left half of the horizontal wire, $\phi+\pi$ in the right half, and $\tilde\phi + \pi$ in the vertical wire.  The `time' $\lambda$ will now vary from $\lambda_e$ to $\lambda_h$, respectively corresponding to the setups of Figs.\ \ref{Tjunction}(e) and (h); the intermediate values $\lambda_f$ and $\lambda_g$ will similarly correspond to Figs.\ \ref{Tjunction}(f) and (g).  For simplicity, as above we evolve the Hamiltonian such that the couplings at each of these `times' are given by $\mu_x,\tilde\mu_y = \mu<0$ and $t_x,\tilde t_y = 0$ in the non-topological regions, while $\mu_x,\tilde\mu_y = 0$ and $t_x,\tilde t_y = t>0$ in the topological regions of the network.  

Suppose that at $\lambda_e$ the Majoranas $\gamma_1$ and $\gamma_2$ reside at sites $X$ and $X'$ of the horizontal wire, respectively; the operators can then be written
\begin{eqnarray}
  \gamma_1(\lambda_e) &=& e^{-i\phi/2} c_X^\dagger + h.c.
  \\
  \gamma_2(\lambda_e) &=& e^{-i(\phi+\pi)/2}c_{X'}^\dagger + h.c.
\end{eqnarray}
[We can safely ignore the other two Majoranas in Fig.\ \ref{Tjunction}(e), since they do not evolve at all during the exchange of $\gamma_{1,2}$ and thus `factor out'; see again Sec.\ \ref{GeneralResults} for a rigorous discussion.  We can even include a coupling between the left and right ends of the horizontal wire if desired to fuse these additional Majoranas.]  
When $\lambda$ increases to $\lambda_f$, we arrive at the setup of Fig.\ \ref{Tjunction}(f) and the Majoranas evolve to
\begin{eqnarray}
  \gamma_1(\lambda_f) &=& \sigma_1\left[e^{-i(\tilde\phi+\pi)/2}\tilde c_1^\dagger + h.c\right].
  \\
  \gamma_2(\lambda_f) &=& \gamma_2(\lambda_e),
\end{eqnarray}
for some sign $\sigma_1$.  Notice that with our modified labeling scheme, the phase factors appearing in the above operators again simply follow from the superconducting phases felt locally by each Majorana.

One can deduce $\sigma_1$ by the usual procedure applied in the previous subsection.  We introduce a deformed Hamiltonian where the superconducting phase in the vertical wire varies spatially from $\tilde\phi+\pi$ at the bottom to $\phi$ at the top, without passing through $\phi+\pi$ to avoid generating spurious zero modes at a $\pi$ junction during this step.  We define the Majoranas $\overline\gamma_{1,2}$ for this deformed Hamiltonian as in Eqs.\ (\ref{phi1}) and (\ref{phi2}); the phases $\phi_{1,2}(\lambda)$ are again precisely the smoothly varying superconducting phases felt locally by $\overline\gamma_{1,2}$.  In particular, as $\overline\gamma_1$ shuttles rightward and then downward, it feels a superconducting phase $\phi_1(\lambda)$ that rotates from $\phi$ at $\lambda_e$ to $\tilde\phi+\pi$ at $\lambda_f$ in an orientation that avoids passing through $\phi+\pi$.  In the example from Fig.\ \ref{BranchCuts}(a), $\phi_1$ rotates from $\phi$ to $\tilde\phi+\pi$ counterclockwise, as the dashed red line indicates.  Because the deformed Hamiltonian and our original Hamiltonian can be smoothly connected throughout this step without closing a gap, we can immediately deduce the sign $\sigma_1$: $\sigma_1 = -1$ if $\phi_1$ crosses the branch cut while $\sigma_1 = +1$ otherwise.   

Returning to the original Hamiltonian, suppose we now transport $\gamma_2$ leftward, generating the configuration of Fig.\ref{Tjunction}(g).  The Majoranas then take the form
\begin{eqnarray}
  \gamma_1(\lambda_g) &=& \gamma_1(\lambda_f)
  \\
  \gamma_2(\lambda_g) &=& \sigma_2\left[e^{-i\phi/2}c_X^\dagger + h.c.\right]
\end{eqnarray}
for some sign $\sigma_2$.  Deducing $\sigma_2$ is trickier than the other signs because midway through this step $\gamma_2$ sits exactly at the trijunction, with all three emanating wire segments being topological.  To analyze this step we begin by introducing a deformed Hamiltonian where the superconducting phase in the horizontal wire varies spatially from from $\phi$ on the left, to $\tilde\phi + \pi$ in the center, to $\phi + \pi$ on the right.  To avoid generating unwanted zero modes during this step, we require that the superconducting phase varies in this fashion without passing through $\tilde\phi$.  As the Majorana $\overline\gamma_2$ moves leftward and approaches the junction, this operator evolves according to Eq.\ (\ref{phi2}) with the local superconducting phase $\phi_2(\lambda)$ varying from $\phi+\pi$ to $\tilde\phi+\pi$ without crossing $\tilde\phi$.  

Eventually $\overline\gamma_2$ sits exactly at the trijunction, and here we need to proceed with care.  Let us denote this point by $\lambda = \lambda_{\rm tri}$.  First, we note that in our deformed Hamiltonian the horizontal wire forms a $\pi$ junction at $\lambda_{\rm tri}$.  Because the vertical wire is also topological, however, no additional zero modes are generated as discussed around Eq.\ (\ref{H123pijunction}).  As we also discussed there, because of the $\pi$ junction the wavefunction for $\overline\gamma_2$ at $\lambda_{\rm tri}$ has no weight on the vertical wire.  We explicitly find that here $\overline\gamma_2$ evolves to
\begin{equation}
  \overline\gamma_2(\lambda_{\rm tri}) = \frac{1}{\sqrt{2}}\left[e^{-i(\tilde\phi+\pi)/2}(-c_N^\dagger + c_{N+2}^\dagger) + h.c.\right].
\end{equation}
(This can be deduced by considering only the sites $N,N+1,N+2$ of the horizontal wire and site $N$ of the vertical wire.)  The minus sign appearing in front of $c_N^\dagger$ is absolutely crucial.  Because of this extra minus sign, as $\overline\gamma_2$ moves off of the trijunction and proceeds leftward, it will subsequently evolve according to
\begin{equation}
  \overline\gamma_2(\lambda>\lambda_{\rm tri}) = -e^{-i\phi_2(\lambda)/2}C_2(\lambda)^\dagger + h.c.,
\end{equation}
where $\phi_2(\lambda)$ corresponds to the superconducting phase felt locally by $\overline\gamma_2$ during the latter half of this step.  Thus $\overline\gamma_2$ feels a local superconducting phase $\phi_2(\lambda)$ which varies smoothly from $\phi+\pi$ at $\lambda_f$ to $\phi$ at $\lambda_g$ in an orientation that passes through $\tilde\phi+\pi$, but additionally picks up an extra minus upon crossing the trijunction.  In the example from Fig.\ \ref{BranchCuts}(b), $\phi_2(\lambda)$ rotates clockwise as shown by the dashed red line.  Using continuity between the deformed and original Hamiltonians, we can now conclude that $\sigma_2 = +1$ if $\phi_2$ crosses the branch cut during this step while $\sigma_2 = -1$ otherwise.  

Finally, let us return to the original Hamiltonian and complete the exchange by transporting $\gamma_1$ up and to the right.  Thus we arrive at the setup of Fig.\ \ref{Tjunction}(h), and the Majoranas become
\begin{eqnarray}
  \gamma_1(\lambda_h) &=& \sigma_1' \sigma_1\left[e^{-i(\phi+\pi)/2}c_{X'}^\dagger + h.c.\right]
  \\
  \gamma_2(\lambda_h) &=& \gamma_2(\lambda_g)
\end{eqnarray}
for some sign $\sigma_1'$ that we can determine by the usual means.  Introduce a deformed Hamiltonian where the superconducting phase in the vertical wire varies from $\tilde\phi+\pi$ at the bottom to $\phi+\pi$ at the top without passing through $\phi$ to avoid additional zero modes.  As the Majorana $\overline\gamma_1$ for this companion problem shuttles up and to the right, it feels a local superconducting phase $\phi_1(\lambda)$ which varies from $\tilde\phi+\pi$ at $\lambda_g$ to $\phi+\pi$ at $\lambda_h$ with precisely this orientation.  In the example of Fig.\ \ref{BranchCuts}(a), $\phi_1$ rotates from $\tilde\phi+\pi$ to $\phi+\pi$ counterclockwise, as the dashed red line shows.  By continuity, we immediately find that $\sigma_1' = -1$ if $\phi_1$ crosses the branch cut during this step while $\sigma_1' = +1$ otherwise.  
  
The final and initial Majorana operators are related by $\gamma_1(\lambda_h) = \sigma_1'\sigma_1\gamma_2(\lambda_e)$ and $\gamma_2(\lambda_h) = \sigma_2\gamma_1(\lambda_e)$.  For the example shown in Fig.\ \ref{BranchCuts} we obtain $\sigma_1 = \sigma_1' = +1$ and $\sigma_2 = -1$, so here $\gamma_1\rightarrow -\gamma_2$ and $\gamma_2 \rightarrow \gamma_1$.  More generally, both $\phi_1$ and $\phi_2$ always rotate by $\pi$ with \emph{opposite} orientations for this type of exchange, and so either $\gamma_1$ and $\gamma_2$ both pick up a minus sign because of the branch cuts or neither do.  The Majorana $\gamma_2$ always acquires an additional minus sign, however, upon crossing the branch cut, so we find as before that
\begin{eqnarray}
  \gamma_1 &\rightarrow& s\gamma_2
  \\
  \gamma_2 &\rightarrow& -s\gamma_1.
\end{eqnarray}
The sign $s$ again depends on $\phi, \tilde\phi$, and how one orients the branch cuts, but is the same for the exchange of Figs.\ \ref{Tjunction}(a)-(d) and that of Figs.\ \ref{Tjunction}(e)-(h).  Thus as one would intuitively expect, both kinds of counterclockwise braids transform the Majoranas in an identical fashion, despite the fact that in one case the Majoranas are initially bridged by a topological region while in the other they initially reside on disconnected topological segments.  This proves our claim in the main text that these exchanges are equivalent.  

\subsubsection{Implications for non-Abelian statistics}

Consider now a network composed of some arbitrary arrangement of trijunctions, such as that of Fig.\ 4(a) from the main text.  The elementary braids of Fig.\ \ref{Tjunction} constitute the basic operations needed to exchange Majorana fermions in this setting.  Putting together all the results obtained so far in this section, exchanging counterclockwise any two Majorana modes $\gamma_i$ and $\gamma_j$ without disturbing any other Majoranas in the system (apart from, perhaps, trivial translations) sends
\begin{eqnarray}
  \gamma_i &\rightarrow& s_{ij}\gamma_j
  \\
  \gamma_j &\rightarrow& -s_{ij} \gamma_i
  \\
  \gamma_k &\rightarrow& \gamma_k ~~~~~~~~~(k\neq i,j).
\end{eqnarray}
The sign $s_{ij}$ depends on the initial sign choices for the operators $\gamma_{i,j}$, the superconducting phases in the system, and how one orients the branch cuts as discussed above.  Once these are specified, $s_{ij}$ can be deduced for any exchange using the simple rules outlined in the previous two subsections.  If $U_{ij}$ denotes the operator that implements this counterclockwise braid, we have
\begin{equation}
  U_{ij} = e^{s_{ij}\pi\gamma_j\gamma_i/4} = \frac{1+s_{ij}\gamma_j\gamma_i}{\sqrt{2}}.
\end{equation}
[One can easily apply the analysis of the previous subsections to the clockwise analogue of the exchanges of Figs.\ \ref{Tjunction}(a)-(d) and Figs.\ \ref{Tjunction}(e)-(h).  This of course leads to the result that the clockwise braid of $\gamma_i$ and $\gamma_j$ is generated by $U_{ij}^{-1} = U_{ij}^\dagger$.]  Non-Abelian statistics for the wire network now follows from the fact that
\begin{equation}
  [U_{ij},U_{jk}] = s_{ij}s_{jk}\gamma_i\gamma_k \neq 0
\end{equation}
for $i \neq k$.  

While trijunctions alone are sufficient to allow for non-Abelian statistics, we note in passing that it is of course possible to consider more general networks featuring some arbitrary number of wires meeting at a junction.  The general results established in Sec.\ \ref{GeneralResults} still apply here, though additional cases can arise beyond those considered in Fig.\ \ref{Tjunction}.  As an example, suppose one fabricated a `+' junction where four wire segments meet at a point.  If the entire junction is topological, then four Majoranas will generically appear.  If we exchange a given pair, which of the braided Majoranas acquires a minus sign can not be immediately deduced from our results above, though this case can be analyzed exactly along the lines of how we studied the exchanges of Fig.\ \ref{Tjunction}.  Our aim is not to be completely exhaustive here, however, so we do not pursue such cases further in this work.  

\subsubsection{Many-body Berry phase calculation for a system with two Majorana fermions I}

Although we have already established non-Abelian statistics in wire networks, in the final parts of this section we will explicitly analyze the evolution of the full many-body ground states under exchange in some tractable cases.  This will serve to not only support our previous analysis, but also enable us to discuss important issues such as the overall Berry phase acquired by the ground states upon braiding Majoranas.  We begin here with the simplest case and return to the initial setup in Fig.\ \ref{Tjunction}(a).  To exchange the Majoranas $\gamma_{1,2}$ as in Fig.\ \ref{Tjunction}(a)-(d), here we will follow the strategy adopted in the main text and keep the Hamiltonian purely real during this exchange (until the very end, when we will allow the Hamiltonian to become complex).  Again, this assumption has the virtue that the wavefunctions can then also be chosen real, so that in spite of their complex evolution the Berry phase accumulated as the Majoranas are transported vanishes identically.  

As in Sec.\ \ref{Fig1exchangeA} we will describe the T-junction by the lattice Hamiltonian in Eq.\ (\ref{Hjunction}).  We will revert back for the rest of the Supplementary Material to the labeling scheme where the superconducting phase is denoted by $\phi$ in the horizontal wire and $\tilde\phi$ in the vertical wire (with respect to the `right' and `up' directions, respectively).   For convenience we will deform the Hamiltonian describing the initial setup of Fig.\ \ref{Tjunction}(a) to the following:
\begin{eqnarray}
  H_i &=& -\tilde \mu \sum_{y = 1}^N \tilde c_y^\dagger \tilde c_y + t\sum_{x = 1}^{2N}(c_x^\dagger+c_x)(c^\dagger_{x+1} -c_{x+1}),
  \label{Hi}
\end{eqnarray}
with $\tilde \mu<0$ and $t>0$.  Here we have set the horizontal wire's superconducting phase to $\phi = 0$ and chemical potential to $\mu_j = 0$, and turned off the pairing and hopping in the vertical wire.  The Hamiltonian then exhibits only real matrix elements as desired.  We graphically denote the initial superconducting phase in the horizontal wire by the rightward-pointing arrow in Fig.\ \ref{Tjunction}(a) (a leftward-pointing arrow would indicate a phase of $\pi$, which would also keep the Hamiltonian purely real).  

The first term in $H_i$ implies that all $\tilde c_y$ fermions in the vertical wire will be absent in the initial ground states, while the second can be recognized as Kitaev's toy model in the special limit where $\mu = 0$, $t = |\Delta|$.  The end Majorana fermions for the horizontal wire take on a particularly simple form in this limit, allowing the initial wavefunctions to be easily obtained.  To do this, we follow Kitaev\cite{1DwiresKitaev} and decompose $c_x$ in terms of Majorana fermions $\gamma_{A/B,x}$ via
\begin{equation}
  c_x = \frac{1}{2}(\gamma_{B,x}+i\gamma_{A,x}),
\end{equation}
which allows the Hamiltonian to be written as
\begin{equation}
  H_i = -\tilde \mu \sum_{y = 1}^N \tilde c_y^\dagger \tilde c_y -it\sum_{x = 1}^{2N}\gamma_{B,x}\gamma_{A,x+1}.
\end{equation}
The zero-energy end Majorana fermions $\gamma_{A,1}$ and $\gamma_{B,2N+1}$ which do not appear in $H$ can be combined into an ordinary zero-energy fermion
\begin{equation}
  d_{\rm end} = \frac{1}{2}(\gamma_{A,1} + i\gamma_{B,2N+1}),
\end{equation}
while the gapped bulk states are captured by operators
\begin{equation}
  d_{x} = \frac{1}{2}(\gamma_{A,x+1} + i\gamma_{B,x}).
\end{equation}
In terms of $d_x$, $H_i$ becomes
\begin{equation}
  H_i = -\tilde \mu \sum_{y = 1}^N \tilde c_y^\dagger \tilde c_y + t\sum_{x = 1}^{2N}(2d_x^\dagger d_x -1).
  \label{Hi2}
\end{equation}
  
The end Majoranas give rise to two degenerate initial ground states whose evolution we are interested in: $|0\rangle_i$ which $d_{\rm end}$ annihilates and $|1\rangle_i = d_{\rm end}^\dagger|0\rangle_i$.  The former can be written $|0\rangle_i = d_{\rm end} \prod_{x = 1}^{2N}d_x |{\rm vac}\rangle$, where $|{\rm vac}\rangle$ denotes the vacuum of $c_x$ and $\tilde c_y$ fermions.  After some algebra, the normalized ground states can be written explicitly as
\begin{eqnarray}
  |0\rangle_i &=& \frac{1}{2^N}\sum_{p=0}^N\sum_{i_1<\cdots<i_{2p+1}}^{2N+1}c_{i_{2p+1}}^\dagger\cdots c_{i_{1}}^\dagger|{\rm vac}\rangle
  \nonumber \\
  |1\rangle_i &=& \frac{1}{2^N}\left[1 + \sum_{p = 1}^{N}\sum_{i_1<\cdots<i_{2p}}^{2N+1}c_{i_{2p}}^\dagger \cdots c_{i_{1}}^\dagger \right]|{\rm vac}\rangle.
  \label{GdSts}
\end{eqnarray}  
Note that we have multiplied $|0\rangle_i$ and $|1\rangle_i$ by overall phase factors to make each wavefunction purely real.  Although the ground states have different fermion parity, both yield the same average particle number
\begin{equation}
  \overline{N} = \frac{2N+1}{2}
\end{equation}
corresponding to half-filling of the horizontal chain.  

Let us now transport the Majorana fermions as outlined in Figs.\ \ref{Tjunction}(a)-(d), keeping the Hamiltonian (and ground state wavefunctions) real and avoiding spurious zero-energy along the way.  For example, $\gamma_1$ can be transported rightward one site by adding the following term to $H_i$,
\begin{equation}
  \delta H = -\lambda \mu c_1^\dagger c_1 -\lambda t (c_1^\dagger + c_1)(c_2^\dagger-c_2)
\end{equation}
(with $\mu<0$) and varying $\lambda$ from 0 to 1.  As usual, as we so transport $\gamma_1$ and $\gamma_2$ we must avoid having two neighboring topological regions whose superconducting phases differ by $\pi$, for in this case a pair of `accidental' zero-energy Majorana modes appears at the junction.  It is therefore useful to employ arrows as shown in Figs.\ \ref{Tjunction}(a)-(d) to signify the sign of the pairing in each topological region.  Two inward or two outward arrows meeting at the junction correspond to a $\pi$ junction and must be avoided.  Figures \ref{Tjunction}(a)-(d) illustrate that in accordance with this simple rule, we can indeed swap the positions of $\gamma_1$ and $\gamma_2$ while keeping the Hamiltonian and wavefunctions purely real, consequently acquiring no Berry phases whatsoever in the process.  However, the arrows and hence the sign of the pairing in the topological region unavoidably reverse, as seen by comparing Figs.\ \ref{Tjunction}(a) and (d).  Thus we have not yet completed an exchange in the usual sense.  

At this stage we have adiabatically evolved the Hamiltonian to
\begin{eqnarray}
    H' &=& -\tilde \mu \sum_{y = 1}^N \tilde c_y^\dagger \tilde c_y - t\sum_{x = 1}^{2N}(c_x^\dagger-c_x)(c^\dagger_{x+1} +c_{x+1}),
\end{eqnarray}
corresponding to $H_i$ with the sign of the pairing reversed, and the wavefunctions to
\begin{eqnarray}
    |0\rangle' &=& \frac{1}{2^N}\sum_{p=0}^N\sum_{i_1<\cdots<i_{2p+1}}^{2N+1}(-1)^p c_{i_{2p+1}}^\dagger\cdots c_{i_{1}}^\dagger|{\rm vac}\rangle
    \nonumber \\
  |1\rangle' &=& \frac{1}{2^N}\left[1 + \sum_{p = 1}^{N}\sum_{i_1<\cdots<i_{2p}}^{2N+1}(-1)^p c_{i_{2p}}^\dagger \cdots c_{i_{1}}^\dagger \right]|{\rm vac}\rangle.
\end{eqnarray}  
[Modulo phase factors, these wavefunctions can be obtained by sending $c_x \rightarrow i c_x$ in Eqs.\ (\ref{GdSts}).]
To complete the exchange, let us now return the Hamiltonian to its original form by adiabatically rotating the superconducting phase in the topological region from $\pi$ back to 0.  The Hamiltonian then involves complex matrix elements, which implies that Berry phases need no longer vanish here.  As we will see, however, the Berry phase contributions for this final step can be easily calculated.  

To this end, consider
\begin{eqnarray}
    H(\lambda) &=& -\tilde \mu \sum_{y = 1}^N \tilde c_y^\dagger \tilde c_y - t\sum_{x = 1}^{2N}(e^{-i\lambda \pi/2}c_x^\dagger-e^{i\lambda \pi/2}c_x)
    \nonumber \\
    &\times& (e^{-i\lambda \pi/2}c^\dagger_{x+1} + e^{i\lambda \pi/2}c_{x+1}).
    \label{Hlambda}
\end{eqnarray}
Upon varying $\lambda$ from 0 to 1, the superconducting phase rotates by $\pi$
such that $H(\lambda = 0) = H'$ and $H(\lambda = 1) = H_i$ as desired.  The ground states of $H(\lambda)$ are
\begin{eqnarray}
    |0(\lambda)\rangle &=& \frac{e^{i\lambda \theta}}{2^N}\sum_{p=0}^N\sum_{i_1<\cdots<i_{2p+1}}^{2N+1}(-1)^p e^{-i\lambda \pi(p+1/2)} 
  \nonumber \\
  &\times& c_{i_{2p+1}}^\dagger\cdots c_{i_{1}}^\dagger|{\rm vac}\rangle 
    \nonumber \\
  |1(\lambda)\rangle &=& \frac{e^{i\lambda \theta}}{2^N}\bigg[1 + \sum_{p = 1}^{N}\sum_{i_1<\cdots<i_{2p}}^{2N+1}(-1)^p e^{-i\lambda \pi p}
    \nonumber \\
    &\times& c_{i_{2p}}^\dagger \cdots c_{i_{1}}^\dagger \bigg]|{\rm vac}\rangle.
  \label{HlambdaGdSts}
\end{eqnarray}  
Importantly, $|0(\lambda = 0)\rangle = |0\rangle'$ and $|1(\lambda = 0)\rangle = |1\rangle'$ so that the wavefunctions evolve smoothly throughout.
Note also that we have inserted an arbitrary phase factor $\theta$ above.  We will select this phase momentarily such that the Berry phase acquired by each wavefunction during this final stage also vanishes.  The outcome of the exchange is then simpler to interpret, since one simply compares the initial states $|0\rangle_i$ and $|1\rangle_i$ with the final states $|0\rangle_f \equiv |0(\lambda = 1)\rangle$ and $|1\rangle_f \equiv |1(\lambda = 1)\rangle$.

Using Eqs.\ (\ref{HlambdaGdSts}), one can now compute the Berry phases; we find
\begin{eqnarray}
  {\rm Im}\int_0^1 d\lambda \langle 0(\lambda)|\partial_\lambda|0(\lambda)\rangle &=&  {\rm Im}\int_0^1 d\lambda \langle 1(\lambda)|\partial_\lambda|1(\lambda)\rangle 
  \nonumber \\
  &=& \theta - \frac{\overline{N}\pi}{2}.
\end{eqnarray}  
(Off-diagonal components such as $\langle 0(\lambda)|\partial_\lambda|1(\lambda)\rangle$ vanish trivially due to the different fermion parity of the ground states.)  This result is quite sensible given that both wavefunctions describe on average $\overline{N}/2$ Cooper pairs whose phase rotates by $\pi$.  We now choose 
\begin{equation}
  \theta = \frac{\overline{N}\pi}{2}
\end{equation}
so that the Berry phases vanish as desired.  Only the explicit relative phases between the initial and final wavefunctions remain.  For $\lambda =1$ the factors of $(-1)^p$ cancel in Eqs.\ (\ref{HlambdaGdSts}), yielding
\begin{eqnarray}
    |0\rangle_f &=& -i e^{i\overline{N}\pi/2}|0\rangle_i
    \nonumber \\
  |1\rangle_f &=& e^{i\overline{N}\pi/2}|1\rangle_i.
  \label{01}
\end{eqnarray}  

Crucially, the ground state $|1\rangle$ acquires an additional phase factor of $i$ relative to $|0\rangle$ under the exchange.  Neglecting an overall phase factor, the unitary operator that generates this relative phase can be written
\begin{equation}
  U_{12} = e^{i\frac{\pi}{4}(2d_{\rm end}^\dagger d_{\rm end}-1)} = e^{\frac{\pi}{4}\gamma_2\gamma_1},
  \label{U12}
\end{equation}
where we have identified $\gamma_1 = \gamma_{A,1}$ and $\gamma_2 = \gamma_{B,2N+1}$.  This coincides with the expression obtained in the main text by somewhat different means, and is identical to the unitary operator generating the exchange of vortices in a spinless $p+ip$ superconductor\cite{Ivanov}.  

Three important comments are warranted here.  First, it is worth emphasizing again that in practice one need not perform any rotation of the superconducting phases to exchange Majoranas or realize non-Abelian statistics, as we have already seen in the preceding subsections.  We analyzed the problem in this way solely because the many-body wavefunctions and Berry phases could be computed very easily in this approach.  In the physical situation appropriate for quantum wires, the effective $p$-wave superconducting phases in the horizontal and vertical wires will differ by $\pi/2$, and to implement the exchange one \emph{only} needs to apply local gate voltages to exchange the Majoranas.  One can also compute the Berry phases in this situation (which we will indeed do momentarily in a simple case), though the calculations are much more complicated.  

Second, while a specific overall phase has been calculated in Eqs.\ (\ref{01}), this phase is certainly non-universal.  It depends both on the precise way in which one exchanges the Majoranas as well details of the Hamiltonian, and is therefore not particularly meaningful in this context.  For instance, had we rotated the superconducting phase from $\pi$ back to 0 before moving $\gamma_1$ all the way to the right in Fig.\ \ref{Tjunction}(d), a different overall phase would emerge.  (As we found earlier, the average number of particles encoded in the wavefunctions when the rotation takes place affects the Berry phase and would be different in this case.)  Furthermore, as we demonstrate below the overall phases depend on the specific form of the Hamiltonian even when the superconducting phases remain fixed; see Eqs.\ (\ref{01b}) and (\ref{4MajoranaStates}).  This result is perhaps not too surprising---the overall phases follow from many-body wavefunctions that encode not only topological information, but also non-universal, high-energy physics.

Third, to obtain the result in Eq.\ (\ref{U12}) we chose in Eq.\ (\ref{Hlambda}) to rotate the superconducting phase counterclockwise from $\pi$ to 0 in the final step of the exchange.  Had we alternatively chosen to rotate the phase in a clockwise fashion, the ground state $|1\rangle$ would pick up a relative phase of $-i$ instead of $+i$ under the exchange compared to $|0\rangle$.  Interesting physics underlies the ambiguity.  To understand this, first note that the cases where one rotates the superconducting phase from $\pi$ to 0 counterclockwise versus clockwise differ by an overall rotation by $2\pi$.  In any superconductor, rotation of the superconducting phase by $2\pi$ effectively changes the sign of all fermion operators; in particular, here such a $2\pi$ rotation sends $\gamma_{1,2}\rightarrow -\gamma_{1,2}$.  Remarkably, exchanging $\gamma_1$ and $\gamma_2$ twice (with the same orientation) while keeping the superconducting phases fixed also sends $\gamma_{1,2}\rightarrow-\gamma_{1,2}$.  In other words, braiding $\gamma_1$ all the way around $\gamma_2$ is equivalent---modulo an overall phase---to a braidless operation wherein the positions of the Majoranas remain fixed but the superconducting phase advances by $2\pi$.  Consequently, if under a counterclockwise exchange $|1\rangle$ picks up a relative phase of $+i$ compared to $|0\rangle$, then subsequently rotating the superconducting phase by $2\pi$ effectively converts this into a clockwise exchange with $|1\rangle$ acquiring a relative phase of $-i$ compared to $|0\rangle$. 

We should emphasize that precisely the same conclusions apply to a 2D spinless $p+ip$ superconductor featuring two vortices.  In that case, however, this observation is less interesting.  With more than two vortices, rotating the overall phase of the 2D $p+ip$ order parameter by $2\pi$ changes the sign of \emph{all} the Majoranas, which is not very useful.  In wire networks, however, by fabricating a series of Josephson junctions along the wires one can in principle wind the superconducting phase only along the region supporting a particular pair of Majoranas, thus effectively implementing pairwise braids in a potentially useful---though not topologically protected--way.  We also note that such ``braidless exchanges'' were recently discussed in a rather different setting by Teo and Kane\cite{TeoKane}.   

\subsubsection{Many-body Berry phase calculation for a system with two Majorana fermions II}

If one transports the Majoranas while leaving the superconducting phases fixed (as one would do in practice), no such ambiguity arises.  To demonstrate this we now consider a four-site model, with the geometry of Fig.\ \ref{LatticeTjunction} with $N = 1$, described by the following Hamiltonian:
\begin{eqnarray}
  H &=& -\mu_1 c_1^\dagger c_1 -\mu_3 c_3^\dagger c_3 -\tilde \mu_1\tilde c_1^\dagger \tilde c_1
  \nonumber \\
  &+& t_{1}(c_1^\dagger + c_1)(c_2^\dagger-c_2) +t_{2}(c_2^\dagger + c_2)(c_3^\dagger-c_3) 
  \nonumber \\
  &+& \tilde t_{1}(e^{i\pi/4}c_2^\dagger + e^{-i\pi/4}c_2)(e^{i\pi/4}\tilde c_1^\dagger - e^{-i\pi/4}\tilde c_1).
  \label{HfixedSCphases}
\end{eqnarray}
As before, $c_{1,2,3}$ correspond to sites on the horizontal chain, while $\tilde c_1$ couples to $c_2$, forming the vertical bond of the T-junction.  The $t_{1,2}$ and $\tilde t_1$ terms represent nearest-neighbor tunneling and pairing with equal amplitude, while $\mu_{1,3},\tilde\mu_1$ denote the chemical potentials (we will never need a chemical potential for $c_2$ and so have excluded such a term above).  Importantly, the superconducting phases have been set to---and will remain fixed at---zero on the horizontal bonds and $-\pi/2$ on the vertical bond, similar to the physical situation for semiconducting wires as discussed in the main text.  

We wish to evolve the coupling constants in Eq.\ (\ref{HfixedSCphases}) to implement an exchange of two Majorana modes as shown in Figs.\ \ref{Tjunction}(a)-(d).  It is simplest to carry out the exchange piecewise in six stages.  We construct the many-body ground states such that they are continuous between steps and compute the Berry phases separately within each step.  Although in principle the wavefunctions can be obtained analytically, the expressions are algebraically very complicated and not terribly illuminating.  Thus we will only outline the calculation here.

In the first step, the couplings evolve according to $\tilde t_{1} = \mu_3 = 0$, $t_{2} = t$, $\tilde \mu_1 = -\mu$, $\mu_1 = -\lambda \mu$, and $t_{1} = t(1-\lambda)$, with $t,\mu>0$ and $\lambda$ varying from 0 to 1.  In the initial configuration with $\lambda = 0$, Majorana fermions $\gamma_1$ and $\gamma_2$ respectively reside on sites 1 and 3 of the horizontal chain, realizing a configuration analogous to Fig.\ \ref{Tjunction}(a).  We define $f = (\gamma_1 + i\gamma_2)/2$ such that the initial ground state $|0\rangle_i$ annihilated by $f$ has odd fermion parity, while the even-parity initial ground state is $|1\rangle_i = f^\dagger |0\rangle_i$.  Ramping up $\lambda\rightarrow 1$ shuttles $\gamma_1$ rightward to site 2.  Since $\tilde t_{1} = 0$ throughout this step, the Hamiltonian is real and the wavefunctions can thus also be chosen real.  The Berry phase therefore vanishes trivially here.

In the second step, the couplings evolve according to $t_{1} = \mu_3 = 0$, $t_{2} = t$, $\mu_1 = -\mu$, $\tilde\mu_1 = -(1-\lambda)\mu$, and $\tilde t_{1} = \lambda t$.  Varying $\lambda$ from 0 to 1 now shuttles $\gamma_1$ downward, realizing a configuration analogous to Fig.\ \ref{Tjunction}(b).  In the third and fourth steps we similarly evolve the couplings to transport $\gamma_2$ from site 3 to site 2, and then from site 2 to site 1, leading to the configuration of Fig.\ \ref{Tjunction}(c).  The fifth step transports $\gamma_1$ upward to site 2.  Note that in steps 2 through 5, the Hamiltonian exhibits complex matrix elements, and non-zero Berry phases thus emerge.  Finally, in step 6 $\gamma_1$ moves rightward to site 3, completing the exchange.  In the last step the Hamiltonian can once again be chosen real, but we explicitly introduce $\lambda$-dependent phase factors into the ground states [as we did in Eqs.\ (\ref{HlambdaGdSts})] such that the Berry phase from step 6 exactly cancels the Berry phase contributions from steps 2 through 5.  All the physics is then contained in the relative phases between the initial and final wavefunctions; we obtain
\begin{eqnarray}
  |0\rangle_f &=& -i e^{-i3\pi/4}|0\rangle_i
  \nonumber \\
  |1\rangle_f &=& e^{-i3\pi/4}|1\rangle_i.
  \label{01b}
\end{eqnarray}
Unlike the case above where we rotated the superconducting phases during the exchange, here the relative phase factors have been determined unambiguously.  Most importantly, the ground state $|1\rangle$ picks up an extra phase factor of $i$ compared to $|0\rangle$ as expected.

\subsubsection{Evolution of many-body ground states under exchange in a toy model exhibiting four Majorana modes}

\begin{figure}
\centering{
  \includegraphics[width=3in]{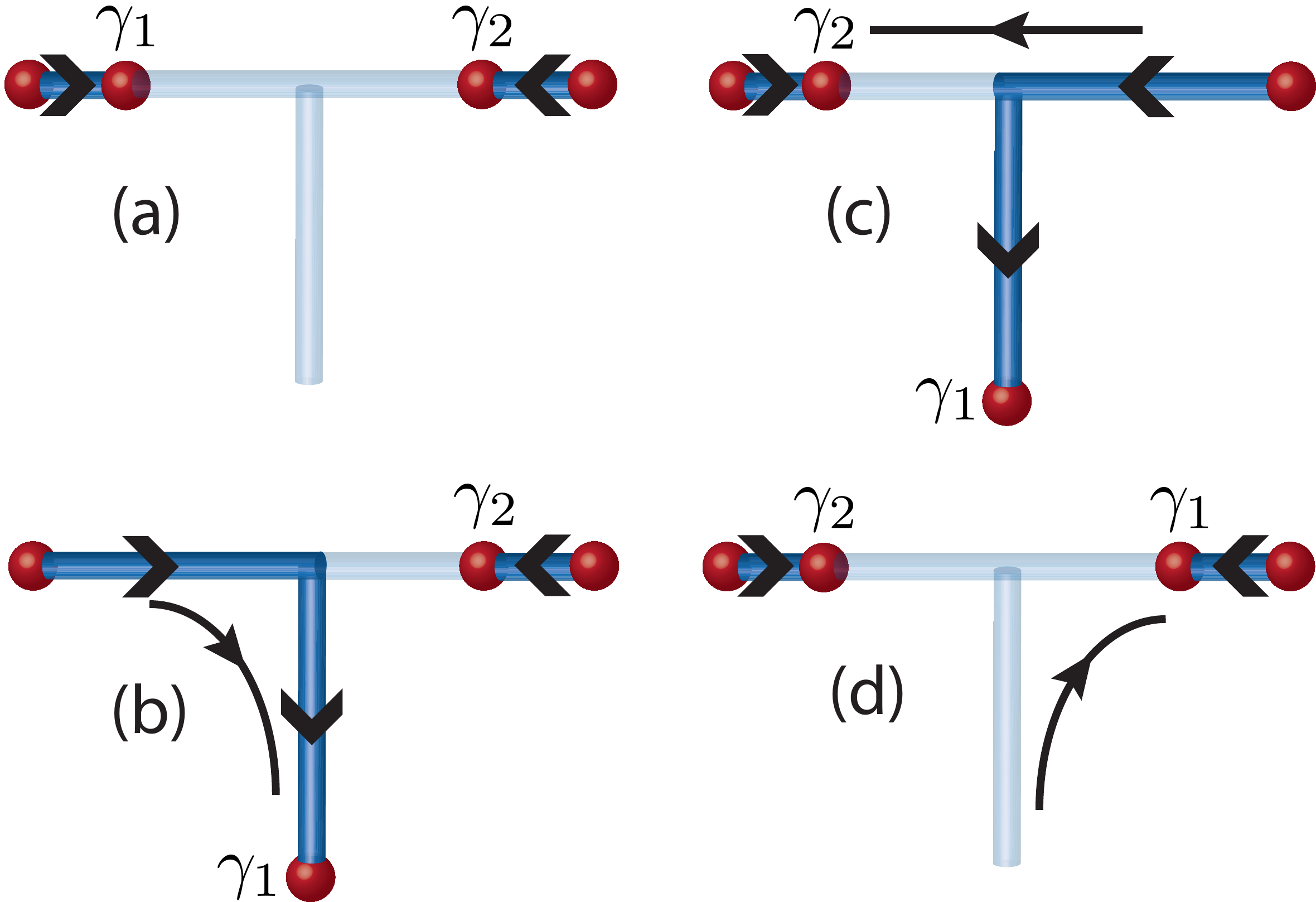}
  \caption{Exchange of two Majorana fermions separated by a non-topological region.  Here we envision transporting the Majoranas while keeping the Hamiltonian purely real.  The superconducting phase in the topological regions can then only take on two values, 0 or $\pi$, which we indicate by arrows above.  Unlike in the exchange of Figs.\ \ref{Tjunction}(a)-(d), here we can exchange Majorana fermions while keeping the Hamiltonian purely real, maintaining the gap, and returning the Hamiltonian back to its original form (\emph{i.e.}, without reversing the sign of the pairing).  As explained in the text, this does \emph{not} mean that the exchange is trivial; indeed, the Majoranas transform just as they do in the braid of Figs.\ \ref{Tjunction}(a)-(d).  }
  \label{Type2exchange}}
\end{figure}

Finally, we will examine how the many-body ground states transform under an exchange of the type outlined in Figs.\ \ref{Tjunction}(e)-(h), where there are four Majoranas present.  We would in particular like to address here the following puzzle.  Suppose we proceed as we did earlier for the exchange of Figs.\ \ref{Tjunction}(a)-(d) and transport $\gamma_{1,2}$ while keeping the Hamiltonian purely real to avoid Berry phase accumulation.  It is once again instructive to view the sign of the pairing in the topological regions with arrows as displayed in Figs.\ \ref{Type2exchange}(a)-(d).  As the figure illustrates, it is now possible to exchange $\gamma_1$ and $\gamma_2$ \emph{without} reversing the sign of the pairing in the process.  In other words, we can keep the Hamiltonian purely real, swap the locations of the Majoranas, and return the Hamiltonian back to its original form---without closing a gap.  One might therefore worry that this type of exchange is trivial, but this is not so.  This can be easily deduced from the perspective of the Majorana operators using the preceding results from this section: upon crossing the trijunction $\gamma_2$ acquires an additional minus sign, and so under the exchange we get the usual result that $\gamma_1\rightarrow s\gamma_2$ and $\gamma_2 \rightarrow - s\gamma_1$, for some sign $s$.  From the perspective of the wavefunctions, how then do the ground states possibly acquire the relative phase factor of $i$ that follows from this transformation?  

The subtlety arises because there are now four Majorana fermions rather than two, and in this case one gets less mileage out of keeping the Hamiltonian real during the exchange.  
To illustrate the point, let $\gamma_{1,2}$ denote the Majoranas we wish to exchange, and $\gamma_{3,4}$ the stationary Majoranas of Figs.\ \ref{Tjunction}(e)-(h).  Defining 
\begin{eqnarray}
  f_A &=& \frac{1}{2}(\gamma_1+i\gamma_2)
  \nonumber \\
  f_B &=& \frac{1}{2}(\gamma_3+i\gamma_4), 
  \label{fAfB}
\end{eqnarray}
we see that there are now two degenerate ground states in each fermion parity sector: $|0,0\rangle$ which both $f_A$ and $f_B$ annihilate, $|1,1\rangle = f_A^\dagger f_B^\dagger |0,0\rangle$, $|1,0\rangle = f_A^\dagger |0,0\rangle$, and $|0,1\rangle = f_B^\dagger |0,0\rangle$.  Reality of the Hamiltonian does \emph{not} imply that these four ground states can each be chosen real.  Indeed, we provide an example below where this is clearly not possible.  Rather, this condition only guarantees reality of specific linear combinations of these ground states, which in general can vary as the exchange takes place.  In other words, the reality condition does not preclude the phases of the above ground states from evolving nontrivially during the exchange (see below for an explicit example).  Drawing conclusions about the exchange from this route therefore requires a more detailed analysis than in the case with only two Majorana fermions.  To remedy this issue, one might be tempted to modify the setup of Figs.\ \ref{Tjunction}(e)-(h) by connecting the horizontal wire into a loop, then fusing $\gamma_3$ and $\gamma_4$ so that only the two Majoranas which we exchange remain.  One will quickly discover, however, that in this case the positions of $\gamma_1$ and $\gamma_2$ can not be swapped while keeping the Hamiltonian real and all other excitations gapped.  Specifically, in the process one necessarily ends up with a configuration similar to Fig.\ \ref{Type2exchange}(c), except with two arrows pointing either into or out of the junction; that is, one can not avoid $\pi$ junctions here.  

\begin{figure}
\centering{
  \includegraphics[width=3in]{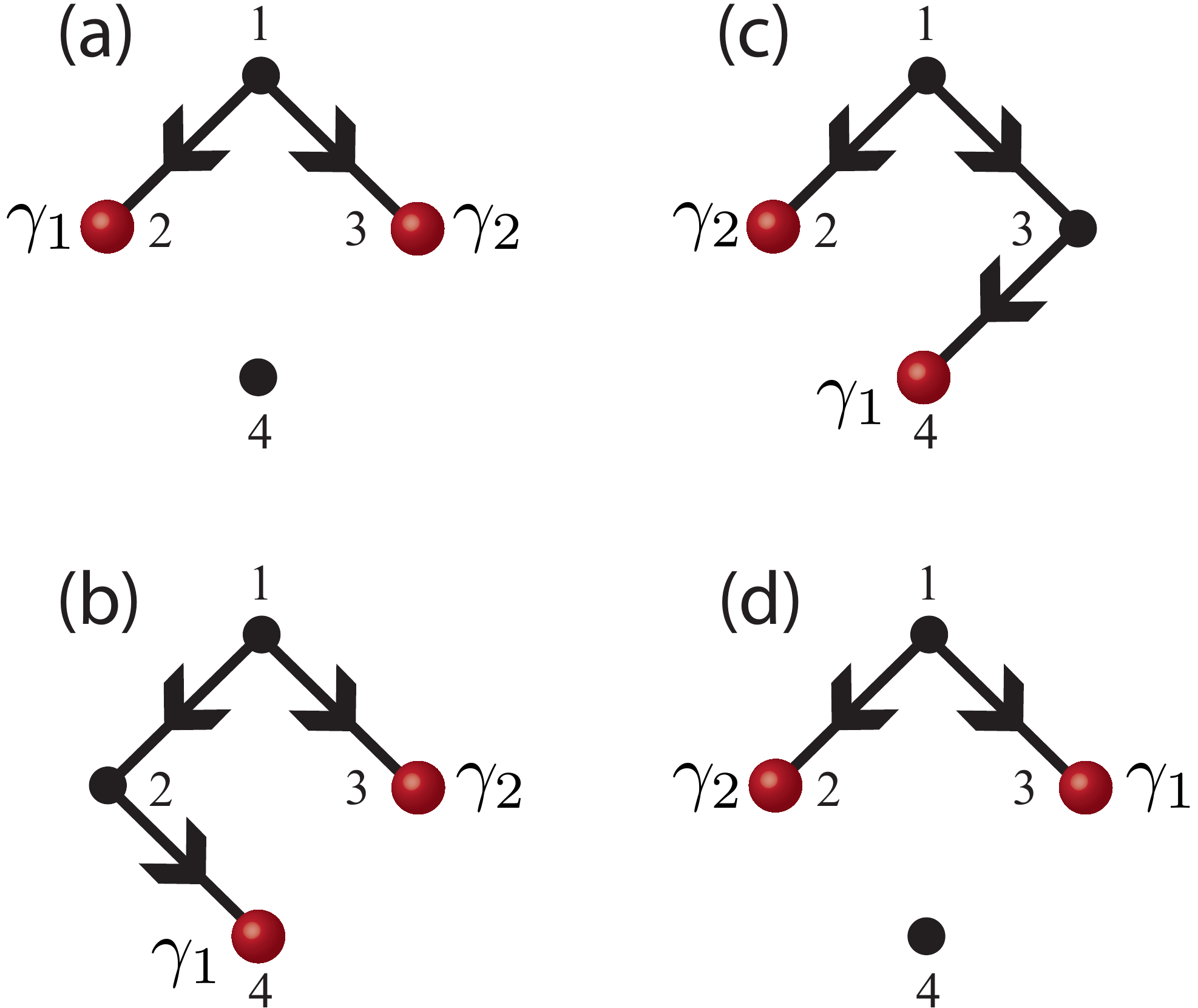}
  \caption{Minimal four-site setup that supports four Majorana modes (only $\gamma_1$ and $\gamma_2$ are shown for clarity).  The Hamiltonian is chosen so as to exchange $\gamma_1$ and $\gamma_2$ as sketched in (a)-(d), mimicking the exchange of Fig.\ \ref{Tjunction}(e)-(h) in a tractable setup.  Solid lines denote bonds with non-zero pairing whose sign is indicated by the arrows.  }
  \label{FourSiteProblem}}
\end{figure}

Let us consider now a toy problem that provides an illustrative minimal setting in which the wavefunctions during such an exchange can be analyzed explicitly.  Specifically, we examine the four-site setup shown in Fig.\ \ref{FourSiteProblem} and described by the following purely real Hamiltonian:
\begin{eqnarray}
  H &=& -\mu c_4^\dagger c_4 + t(c_1^\dagger + c_1)[(c_2^\dagger -c_2) +(c_3^\dagger-c_3)]
  \nonumber \\
  &+& [t_{24}(c_2^\dagger + c_2) + t_{34}(c_3^\dagger+c_3)](c_4^\dagger-c_4),
  \label{H4site}
\end{eqnarray}
with $t>0$, $\mu\leq 0$, and $t_{24},t_{34}\geq 0$.  In spite of the small number of sites, this Hamiltonian supports four zero-energy Majorana modes for any values of $\mu$, $t_{24}$, and $t_{34}$.  To get intuition here, it is useful to think of site 1 as forming a $\pi$ junction between sites 2 and 3.  This gives rise to two Majorana modes which are independent of the parameters appearing in Eq.\ (\ref{H4site}).  One of these, $\gamma_3$, resides at site 1:
\begin{equation}
  \gamma_3 = i(c_1^\dagger-c_1).
\end{equation}
The other, $\gamma_4$, resides on sites 2 and 3:
\begin{equation}
  \gamma_4 = \frac{i}{\sqrt{2}}(c_2^\dagger-c_2-c_3^\dagger+c_3).
\end{equation}
The locations of the second pair of Majoranas, $\gamma_1$ and $\gamma_2$, depend on $\mu$, $t_{24}$, and $t_{34}$.  We will vary these parameters so as to carry out an exchange of $\gamma_1$ and $\gamma_2$ in a manner that is analogous to the exchange of Figs.\ \ref{Tjunction}(e)-(h).  

To help establish a connection between the setup of Fig.\ \ref{Tjunction}(e) and the present four site-problem, imagine first forming a loop out of the horizontal wire so that the two topological regions connect.  The outer Majoranas of Fig.\ \ref{Tjunction}(e)---which are analogous to $\gamma_{3,4}$ in our setup---can then be generated by forming a $\pi$ junction in the topological region.  The four-site problem shrinks this $\pi$ junction to the smallest possible size.  The other two Majoranas $\gamma_{1,2}$ will initially be separated by an unpaired region, similar to the non-topological segments connecting $\gamma_{1,2}$ in Fig.\ \ref{Tjunction}(e).  We carry out their exchange piecewise in three stages to reduce the algebraic complexity of the problem.  Care will be taken to ensure that the wavefunctions and operators defined below evolve continuously in between each of these stages.  

(I) In the first stage, we evolve the Hamiltonian by taking
\begin{eqnarray}
  \mu &=& (1-\lambda)\mu_4
  \nonumber \\
  t_{24} &=& \lambda t
  \\
  t_{34} &=& 0
  \nonumber
\end{eqnarray}
(with $\mu_4<0$) and varying $\lambda$ from 0 to 1.  Initially when $\lambda = 0$, $\gamma_1$ and $\gamma_2$ are situated at sites 2 and 3, respectively, as Fig.\ \ref{FourSiteProblem}(a) illustrates.  Ramping up $\lambda$ to 1 shuttles $\gamma_1$ to site 4, leading to the configuration of Fig.\ \ref{FourSiteProblem}(b).  More precisely, $\gamma_{1,2}$ are given by
\begin{eqnarray}
  \gamma_1^I &=& \frac{1}{\alpha_{24}}[-\mu(c_2^\dagger + c_2)+2t_{24}(c_4^\dagger + c_4)]
  \\
  \gamma_2^I &=& c_3^\dagger + c_3,
\end{eqnarray}
where we have defined $\alpha_{24} = \sqrt{4t_{24}^2+\mu^2}$.  The finite-energy fermion operators which annihilate the ground states are
\begin{eqnarray}
  d_A^I &=& \frac{1}{2\sqrt{2}}[\sqrt{2}(c_1^\dagger+c_1)+(c_2^\dagger-c_2)+(c_3^\dagger-c_3)]
  \\
  d_B^I &=& \frac{1}{2\alpha_{24}}[2t_{24}(c_2^\dagger+c_2)+(\mu+\alpha_{24})c_4^\dagger+(\mu - \alpha_{24})c_4].
  \nonumber \\
\end{eqnarray}

Now define $f_A^I$ and $f_B$ analogously to Eqs.\ (\ref{fAfB}).  Suppressing the $\lambda$ dependence for notational simplicity, the four degenerate ground states are then 
\begin{eqnarray}
  |0,0\rangle_I && 
  \nonumber \\
  |1,1\rangle_I &=& f_A^{I\dagger} f_B^\dagger |0,0\rangle_I
  \\
  |1,0\rangle_I &=& f_A^{I\dagger} |0,0\rangle_I
  \nonumber \\
  |0,1\rangle_I &=& f_B^\dagger|0,0\rangle_I,
  \nonumber
\end{eqnarray}  
where $|0,0\rangle_I$ is annihilated by $f_A^I,f_B, d_A^I$, and $d_B^I$.  With the above definitions and some time to carry out the algebra, one can obtain these ground states for arbitrary $\lambda$.
When $\lambda = 0$ leading to the initial configuration shown in Fig.\ \ref{FourSiteProblem}(a), the wavefunctions are
\begin{eqnarray}
  |0,0\rangle_i &=& \frac{1}{2}[-i-e^{i\frac{\pi}{4}}c_2^\dagger c_1^\dagger + e^{-i\frac{\pi}{4}}c_3^\dagger c_1^\dagger + c_3^\dagger c_2^\dagger]|{\rm vac}\rangle
  \nonumber \\
  |1,1\rangle_i &=& \frac{1}{2}[e^{-i\frac{\pi}{4}}+c_2^\dagger c_1^\dagger -i c_3^\dagger c_1^\dagger -e^{i\frac{\pi}{4}}c_3^\dagger c_2^\dagger]|{\rm vac}\rangle
  \\
  |1,0\rangle_i &=& \frac{1}{2}[-e^{i\frac{\pi}{4}}c_1^\dagger -i c_2^\dagger - c_3^\dagger-e^{-i\frac{\pi}{4}} c_3^\dagger c_2^\dagger c_1^\dagger]|{\rm vac}\rangle
  \nonumber \\
  |0,1\rangle_i &=& \frac{1}{2}[c_1^\dagger + e^{-i\frac{\pi}{4}} c_2^\dagger+ e^{i\frac{\pi}{4}} c_3^\dagger +i c_3^\dagger c_2^\dagger c_1^\dagger]|{\rm vac}\rangle.
  \nonumber
\end{eqnarray}
Clearly none of these can be made real by introducing overall phase factors (though one can readily verify that a purely real basis does exist by considering linear combinations of these states).  
When $\lambda \rightarrow 1$ and we arrive at the configuration shown in Fig.\ \ref{FourSiteProblem}(b), the wavefunctions evolve to
\begin{eqnarray}
  |0,0\rangle_b &=& \frac{1}{2\sqrt{2}}[-i -e^{i\frac{\pi}{4}}c_2^\dagger c_1^\dagger + e^{-i\frac{\pi}{4}}c_3^\dagger c_1^\dagger -e^{i\frac{\pi}{4}}c_4^\dagger c_1^\dagger
  \nonumber \\
  &+& c_3^\dagger c_2^\dagger -i c_4^\dagger c_2^\dagger -c_4^\dagger c_3^\dagger-e^{-i\frac{\pi}{4}}c_4^\dagger c_3^\dagger c_2^\dagger c_1^\dagger]|{\rm vac}\rangle
  \nonumber \\
  |1,1\rangle_b &=& \frac{1}{2\sqrt{2}}[e^{-i\frac{\pi}{4}} +c_2^\dagger c_1^\dagger -i c_3^\dagger c_1^\dagger + c_4^\dagger c_1^\dagger
  - e^{i\frac{\pi}{4}} c_3^\dagger c_2^\dagger 
  \nonumber \\
  &+& e^{-i\frac{\pi}{4}} c_4^\dagger c_2^\dagger +e^{i\frac{\pi}{4}}c_4^\dagger c_3^\dagger+ i c_4^\dagger c_3^\dagger c_2^\dagger c_1^\dagger]|{\rm vac}\rangle
  \label{b} \\
  |1,0\rangle_b &=& \frac{1}{2\sqrt{2}}[-e^{i\frac{\pi}{4}}c_1^\dagger-i c_2^\dagger-c_3^\dagger-i c_4^\dagger-e^{-i\frac{\pi}{4}}c_3^\dagger c_2^\dagger c_1^\dagger
  \nonumber \\
  &-&e^{i\frac{\pi}{4}}c_4^\dagger c_2^\dagger c_1^\dagger+e^{-i\frac{\pi}{4}}c_4^\dagger c_3^\dagger c_1^\dagger+c_4^\dagger c_3^\dagger c_2^\dagger]|{\rm vac}\rangle
  \nonumber \\
  |0,1\rangle_b &=& \frac{1}{2\sqrt{2}}[c_1^\dagger + e^{-i\frac{\pi}{4}}c_2^\dagger +e^{i\frac{\pi}{4}}c_3^\dagger + e^{-i\frac{\pi}{4}}c_4^\dagger
  \nonumber \\
  &+& ic_3^\dagger c_2^\dagger c_1^\dagger+c_4^\dagger c_2^\dagger c_1^\dagger -i c_4^\dagger c_3^\dagger c_1^\dagger-e^{i\frac{\pi}{4}}c_4^\dagger c_3^\dagger c_2^\dagger]|{\rm vac}\rangle.
  \nonumber
\end{eqnarray}

(II) For the second stage of the exchange, we evolve the Hamiltonian according to
\begin{eqnarray}
  \mu &=& 0
  \nonumber \\
  t_{24} &=& (1-\lambda)t
  \\
  t_{34} &=& \lambda t.
  \nonumber
\end{eqnarray}
Here varying $\lambda$ from 0 to 1 leaves $\gamma_1$ unchanged but adiabatically transports $\gamma_2$ from site 3 to site 2, leading to the configuration of Fig.\ \ref{FourSiteProblem}(c).  Defining $\beta = \sqrt{t_{34}^2 + t_{24}^2}$, the Majorana fermion operators at this stage obey
\begin{eqnarray}
  \gamma_1^{II} &=& c_4^\dagger + c_4
  \\
  \gamma_2^{II} &=& \frac{1}{\beta}[-t_{34}(c_2^\dagger + c_2) + t_{24}(c_3^\dagger + c_3)],
\end{eqnarray}
while the gapped quasiparticle operators are
\begin{eqnarray}
  d_A^{II} &=& d_A^I
  \\
  d_B^{II} &=& \frac{1}{2\beta}[t_{24}(c_2^\dagger+c_2)+t_{34}(c_3^\dagger+c_3)+\beta(c_4^\dagger-c_4)].
\end{eqnarray}
(Note that there is some freedom for how one implements this step.  The expression for $\gamma_2^{II}$ does not depend on the specific parametrization of $t_{24}$ and $t_{34}$ above.  If desired, one can first turn on $t_{34}$ while keeping $t_{24}$ fixed, and then turn off $t_{24}$.  The end result in either case is the same.)  The wavefunctions can again be obtained for arbitrary $\lambda$ after some tedious algebra.  In particular, when $\lambda\rightarrow 1$ bringing the system to the setup of Fig.\ \ref{FourSiteProblem}(c), the wavefunctions evolve to
\begin{eqnarray}
  |0,0\rangle_c &=& \frac{1}{2\sqrt{2}}[-i -e^{i\frac{\pi}{4}}c_2^\dagger c_1^\dagger + e^{-i\frac{\pi}{4}}c_3^\dagger c_1^\dagger + e^{-i\frac{\pi}{4}}c_4^\dagger c_1^\dagger
  \nonumber \\
  &+& c_3^\dagger c_2^\dagger + c_4^\dagger c_2^\dagger -i c_4^\dagger c_3^\dagger -e^{i\frac{\pi}{4}}c_4^\dagger c_3^\dagger c_2^\dagger c_1^\dagger]|{\rm vac}\rangle
  \nonumber \\
  |1,1\rangle_c &=& \frac{1}{2\sqrt{2}}[e^{i\frac{\pi}{4}}+i c_2^\dagger c_1^\dagger + c_3^\dagger c_1^\dagger + c_4^\dagger c_1^\dagger +e^{-i\frac{\pi}{4}}c_3^\dagger c_2^\dagger
  \nonumber \\
  &+& e^{-i\frac{\pi}{4}}c_4^\dagger c_2^\dagger +e^{i\frac{\pi}{4}}c_4^\dagger c_3^\dagger + i c_4^\dagger c_3^\dagger c_2^\dagger c_1^\dagger]|{\rm vac}\rangle
  \label{c} \\
  |1,0\rangle_c &=& \frac{1}{2\sqrt{2}}[e^{-i\frac{\pi}{4}}c_1^\dagger + c_2^\dagger -i c_3^\dagger -i c_4^\dagger -e^{i\frac{\pi}{4}}c_3^\dagger c_2^\dagger c_1^\dagger
  \nonumber \\
  &-& e^{i\frac{\pi}{4}}c_4^\dagger c_2^\dagger c_1^\dagger + e^{-i\frac{\pi}{4}}c_4^\dagger c_3^\dagger c_1^\dagger + c_4^\dagger c_3^\dagger c_2^\dagger]|{\rm vac}\rangle
  \nonumber \\
  |0,1\rangle_c &=& \frac{1}{2\sqrt{2}}[c_1^\dagger + e^{-i\frac{\pi}{4}}c_2^\dagger + e^{i\frac{\pi}{4}}c_3^\dagger + e^{i\frac{\pi}{4}}c_4^\dagger
  \nonumber \\
  &+& i c_3^\dagger c_2^\dagger c_1^\dagger + i c_4^\dagger c_2^\dagger c_1^\dagger +c_4^\dagger c_3^\dagger c_1^\dagger +e^{-i\frac{\pi}{4}}c_4^\dagger c_3^\dagger c_2^\dagger]|{\rm vac}\rangle.
  \nonumber
\end{eqnarray}
Notice how the phase factors in the wavefunctions evolve nontrivially in passing from Eqs.\ (\ref{b}) to (\ref{c}), despite the reality of the Hamiltonian.  

(III) To conclude the exchange, we now choose
\begin{eqnarray}
  \mu &=& \lambda \mu_4
  \nonumber \\
  t_{24} &=& 0
  \\
  t_{34} &=& (1-\lambda)t
  \nonumber
\end{eqnarray}
and again vary $\lambda$ from 0 to 1.  In this final step, $\gamma_2$ remains unchanged while $\gamma_1$ moves adiabatically to site 3 as in Fig.\ \ref{FourSiteProblem}(d).  The Majorana operators now obey
\begin{eqnarray}
  \gamma_1^{III} &=& \frac{1}{\alpha_{34}}[-\mu(c_3^\dagger+c_3)+2t_{34}(c_4^\dagger+c_4)]
  \\
  \gamma_2^{III} &=& -(c_2^\dagger+c_2),
\end{eqnarray}
with $\alpha_{34} = \sqrt{4t_{34}^2 + \mu^2}$, and the gapped quasiparticle operators are
\begin{eqnarray}
  d_A^{III} &=& d_A^I
  \\
  d_B^{III} &=& \frac{1}{2\alpha_{34}}[2t_{34}(c_3^\dagger+c_3) + (\mu+\alpha_{34})c_4^\dagger+(\mu-\alpha_{34})c_4].
  \nonumber \\
\end{eqnarray}
Computing the ground states as before, we obtain the expected result that the final and initial ground states are related by
\begin{eqnarray}
  |0,0\rangle_f &=& |0,0\rangle_i
  \nonumber \\
  |1,1\rangle_f &=& i|1,1\rangle_i
  \label{4MajoranaStates} \\
  |1,0\rangle_f &=& i|1,0\rangle_i
  \nonumber \\
  |0,1\rangle_f &=& |0,1\rangle_i.
  \nonumber
\end{eqnarray}
That is, the ground states with an $f_A = (\gamma_1+i\gamma_2)/2$ fermion present acquire a phase factor of $i$ under the exchange.  Up to an overall non-universal phase factor, this transformation is generated by the unitary operator $U_{12} = e^{\pi\gamma_2\gamma_1/4}$, as obtained earlier.  

To close this section, we remark that one may object that in obtaining this result we have simply compared the initial and final states.  Since the above wavefunctions are not real, one may in particular ask whether the exchange is tainted by Berry phases.  It is not---it is always possible to simply change to a real basis by suitably superposing these wavefunctions, and in such a basis the absence of Berry phases is manifest.  The exchange indeed is governed solely by the difference between initial and final states.  

\subsection{Derivation of the fractional Josephson effect in a simple model}

\begin{figure}
\centering{
  \includegraphics[width=3in]{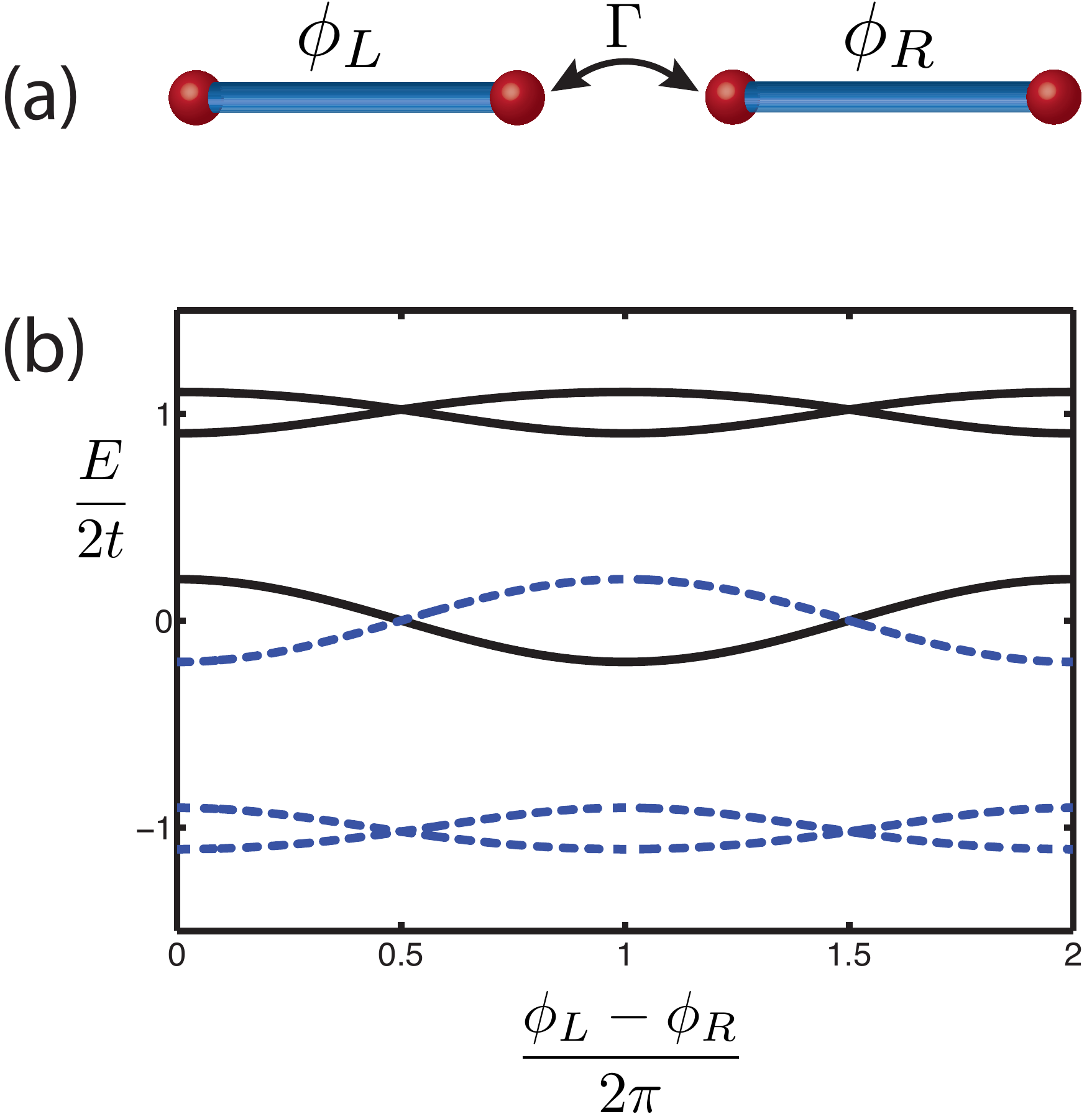}
  \caption{(a) Schematic of the Josephson junction formed by two topological wires with $p$-wave superconducting phases $\phi_{L/R}$.  The wires couple at the junction through an electron tunneling term with strength $\Gamma$.  (b) Bogoliubov-de Gennes spectrum as a function of $(\phi_L-\phi_R)/(2\pi)$ for the effective Hamiltonian in Eq.\ (\ref{Heff2}) chosen to describe the junction.  Only the solid lines denote physically distinct states.  The states centered around $E/(2t) = 1$ represent ordinary bulk quasiparticles, while the state near zero energy represents the quasiparticle formed when the two end Majoranas at the junction fuse.  The energy and hence Josephson current corresponding to the latter exhibit $4\pi$ periodicity in $\phi_L-\phi_R$.  The ordinary bulk quasiparticle states, however, contribute only to the usual $2\pi$-periodic Josephson effect.}
  \label{JosephsonFig}}
\end{figure}

For pedagogical purposes, we will review here the `fractional Josephson effect' originally predicted by Kitaev\cite{1DwiresKitaev} and discussed by other authors in the context of 1D wires\cite{1DwiresLutchyn,1DwiresOreg} and other topological systems\cite{Kwon,Kwon2, FuKane,MajoranaQSHedge}.  We will examine this effect in a minimal setup where all calculations can be explicitly carried out, although the qualitative aspects of the physics are more universal.  Consider two topological superconducting wires forming a Josephson junction as shown schematically in Fig.\ \ref{JosephsonFig}(a).  The phases of the $p$-wave order parameters are taken to be $\phi_{L/R}$ in the left/right wires, which are coupled by a weak (compared to the gap in each wire) electron tunneling term at the junction.  The full Hamiltonian reads
\begin{equation}
  H = H_L + H_R + H_\Gamma,
\end{equation}
where $H_{L/R}$ describe the left/right regions and $H_\Gamma$ represents the electron tunneling term coupling the wires.  For computational simplicity, we model the left and right regions as $N$-site chains described by Kitaev's toy model with $\mu = 0$ and $t = |\Delta|$.  In this case we have
\begin{eqnarray}
  H_\alpha &=& t\sum_{x = 1}^{N-1}(e^{-i\phi_\alpha/2}c_{\alpha,x}^\dagger+e^{i\phi_\alpha/2}c_{\alpha,x})
  \nonumber \\
  &\times& (e^{-i\phi_\alpha/2}c_{\alpha,x+1}^\dagger-e^{i\phi_\alpha/2}c_{\alpha,x+1}),
\end{eqnarray}
for $\alpha = L/R$, along with a tunneling term
\begin{equation}
  H_\Gamma = -\Gamma(c_{L,N}^\dagger c_{R,1} + h.c.).
\end{equation}
When $\Gamma = 0$, two Majorana fermions reside at the junction; turning on $\Gamma \neq 0$ generally fuses these to an ordinary finite-energy quasiparticle state.  
We wish to compute the zero-bias current flowing across the junction,
\begin{equation}
  I = -\frac{e \Gamma}{\hbar} \langle ic_{L,N}^\dagger c_{R,1} + h.c.\rangle,
\end{equation}
in the ground state as well as the excited state where this quasiparticle state is occupied. 

We proceed by first diagonalizing $H_{L/R}$ in the usual way.  Writing $c_{\alpha,x} = e^{-i\phi_\alpha/2}(\gamma^\alpha_{B,x}+i \gamma_{A,x}^\alpha)/2$ and then defining $d_{\alpha,x} = (\gamma^\alpha_{A,x+1}+i \gamma^\alpha_{B,x})/2$, one obtains 
\begin{equation}
  H_{\alpha} = t\sum_{x = 1}^{N-1}(2d_{\alpha,x}^\dagger d_{\alpha,x}-1).
\end{equation}
It is useful to group the end Majorana fermions residing at the junction into an ordinary fermion operator via
\begin{equation}
  d_{\rm end} = \frac{1}{2}(\gamma^R_{A,1}+i \gamma_{B,N}^L).
\end{equation}
The tunneling term, which we will treat as a perturbation, can then be written
\begin{eqnarray}
  H_\Gamma &=& \frac{\Gamma}{2}\{ C[d_{R,1}^\dagger(d_{L,N-1}^\dagger+d_{L,N-1})+h.c.]
  \nonumber \\
  &+&S[i d_{\rm end}^\dagger(d_{R,1}^\dagger-d_{R,1}+d_{L,N-1}^\dagger+d_{L,N-1}) + h.c.]
  \nonumber \\
  &+& C(2d_{\rm end}^\dagger d_{\rm end}-1)\}
\end{eqnarray}
with 
\begin{eqnarray}
  C &=& \cos(\Delta\phi/2)
  \\
  S &=& \sin(\Delta\phi/2)
  \\
  \Delta \phi &=& \phi_L-\phi_R.
\end{eqnarray}
Rewriting the expression for the current in this basis, one obtains the familiar relation
\begin{equation}
  I = \frac{2e}{\hbar}\frac{d\langle H_\Gamma\rangle}{d\Delta\phi}.
  \label{current}
\end{equation}

Notice that the fermion operators $d_{L,1\ldots N-2}$ and $d_{R,2\ldots N-1}$ essentially drop out from the problem---the full Hamiltonian separately conserves the fermion number for each of these states and they do not contribute to the Josephson current.  Thus for the purposes of evaluating the current, the problem maps onto a simpler Hamiltonian involving only $d_{\rm end}$, $d_{L,N-1}$, and $d_{R,1}$.  In terms of $d_A = (d_{L,N-1}+d_{R,1})/\sqrt{2}$ and $d_B = (d_{L,N-1}-d_{R,1})/\sqrt{2}$, this effective Hamiltonian becomes
\begin{eqnarray}
  H_{\rm eff} &=& t[(2d_A^\dagger d_A-1)+(2d_B^\dagger d_B-1)] + H_{\Gamma},
  \label{Heff2}
\end{eqnarray}
where now
\begin{eqnarray}
  H_{\Gamma} &=& \frac{\Gamma}{2}\{C[(2d_{\rm end}^\dagger d_{\rm end}-1)+ (d_A^\dagger d_A -d_B^\dagger d_B)]
  \nonumber \\
  &+& C(d_A^\dagger d_B^\dagger + h.c.) + \sqrt{2}S[i d_{\rm end}^\dagger(d_A^\dagger + d_B) + h.c.]\}.
  \nonumber \\
  \label{HGamma}
\end{eqnarray}
Applying degenerate perturbation theory to obtain the energies of the $d_{\rm end}$, $d_A$, and $d_B$ fermions to $O[(\Gamma/t)^2]$, we obtain
\begin{eqnarray}
  H_{\rm eff} &\approx& E_A \left(f_A^\dagger f_A-\frac{1}{2}\right) + E_B \left(f_B^\dagger f_B-\frac{1}{2}\right)
  \nonumber \\
  &+& E_{\rm end} \left(f_{\rm end}^\dagger f_{\rm end}-\frac{1}{2}\right)
  \label{HeffDiagonal}
\end{eqnarray}
The operators $f_{A/B/{\rm end}}$ correspond to states that evolve from $d_{A/B/{\rm end}}$ due to the tunneling perturbation.  Their energies to the desired order are
\begin{eqnarray}
  E_{A/B} &=& 2t \pm \frac{\Gamma}{2}\cos(\Delta\phi/2)
  + \frac{\Gamma^2}{32t}[5-3\cos\Delta\phi]
  \label{EAB}
  \\
  E_{\rm end} &=& \Gamma\cos(\Delta\phi/2).
\end{eqnarray}

We can now evaluate the Josephson current in the ground state, as well as the excited state where the finite-energy quasiparticle formed from the fused Majoranas is occupied.  Equations (\ref{current}) and (\ref{HeffDiagonal}), along with the above energies, yield
\begin{eqnarray}
  I_{\pm} &=& \pm \frac{e\Gamma}{2\hbar}\sin(\Delta\phi/2)-\frac{3 e\Gamma^2}{16\hbar t}\sin\Delta\phi,
\end{eqnarray}
where the $+/-$ sign corresponds to the current obtained when the $f_{\rm end}$ fermion is unoccupied/occupied.  The second term represents the standard Josephson current that is $2\pi$ periodic in $\Delta\phi$.  This contribution reflects Cooper-pair tunneling and thus arises at second-order in perturbation theory.  More interestingly, the first term exhibits $4\pi$ periodicity and has a topological origin since it arises solely from the Majoranas fused at the junction.  This contribution reflects a \emph{first-order} process corresponding to single-electron tunneling, which is possible at zero bias because the Majoranas form a zero-energy state at the junction when $\Gamma = 0$.  

It is interesting to observe from Eq.\ (\ref{EAB}) that the $f_{A/B}$ fermions also pick up a first order correction to their energy from $\Gamma$.  Thus one can view each of these states as individually contributing both $2\pi$- \emph{and} $4\pi$-periodic Josephson currents.  Their $4\pi$-periodic contributions exactly cancel one another, however, so that only the fused Majoranas contribute to this effect.  Mathematically, this can be understood from the particle/hole-symmetric spectrum of the Bogoliubov-de Gennes Hamiltonian in Eq.\ (\ref{Heff2}).  This is plotted versus $\Delta\phi/(2\pi)$ in Fig.\ \ref{JosephsonFig}(b).  Here only the solid lines denote physically distinct states since those with energy $E$ and $-E$ are not independent.  As one turns on the tunneling strength $\Gamma$ from 0, the ordinary fermionic states that begin at energy $2t$ split with opposite sign at first order, and only yield a net change in energy at second order in $\Gamma/t$.  Thus they contribute only to the usual $2\pi$-periodic Josephson current.  The state of affairs for the end state which begins at zero energy is very different---its energy also shifts at first order, but its `partner' which shifts in the opposite direction does not represent a physically distinct state.  It therefore produces an observable $4\pi$-periodic Josephson current.  

Finally, it is useful to ask whether the crossings in the spectrum of Fig.\ \ref{JosephsonFig}(b) at $\Delta\phi = \pi$ are stable.  In the case of the ordinary $f_{A,B}$ quasiparticle states, they are certainly not.  For instance, adding a weak superconducting pairing between $c_{L,N}$ and $c_{R,1}$ at the junction lifts the crossings near $E = \pm 2t$ in the figure.  When this happens, there is no sense in which the bulk quasiparticle states even individually contribute a $4\pi$-periodic current.  The crossing at $E = 0$, however, \emph{is} stable provided the Majoranas at the outer ends of the wires do not overlap with those at the junction\cite{1DwiresKitaev}.  (The location of the crossing\cite{1DwiresLutchyn} though need not occur exactly at $\Delta\phi = \pi$).  This can be understood as follows.  As long the as occupation number of the fermion corresponding to the outer Majorana end states remains fixed, the ground states at $\Delta\phi = 0$ and $\Delta\phi = 2\pi$ have different fermion parity.  In the former case $f_{\rm end}^\dagger f_{\rm end} = 0$ in the ground state while in the latter $f_{\rm end}^\dagger f_{\rm end} = 1$.  If this crossing could be removed, then one would be able to adiabatically evolve $\Delta\phi$ from 0 to $2\pi$ while remaining in the ground state, but this can not happen unless the outer end Majoranas transfer a fermion to the junction.  It is useful to keep this in mind when considering the left path in Fig.\ 4(b) of the main text, where $\gamma_2$ crosses the junction.  At fixed phase $\Delta\varphi^i\neq \pi$, the ground state is \emph{always} accessible here, precisely because $\gamma_2$ overlaps with the fused Majoranas $\gamma_{3,4}$ at the junction during this process.  However, when $\gamma_2$ resides at the far right end of the wire, then (neglecting residual overlap between the Majoranas) the ground state will no longer accessible when the phase difference changes by $2\pi$.  This explains why the fractional Josephson current in Eq.\ (9) of the main text exhibits $2\pi$ periodicity in the initial phase difference $\Delta \varphi^i$, but $4\pi$ periodicity in $\Delta\varphi$.

\acknowledgments{We have benefited greatly from stimulating conversations with Parsa Bonderson, Sankar Das Sarma, Lukasz Fidkowski, Erik Henriksen, Alexei Kitaev, Patrick Lee, Xiaoliang Qi, and Ady Stern.  We also gratefully acknowledge support from the Lee A.\ DuBridge Foundation, ISF, BSF, DIP, and SPP 1285 grants, Packard and Sloan fellowships, the Institute for Quantum Information under NSF grants PHY-0456720 and PHY-0803371, and the National Science Foundation through grant DMR-0529399.}

%\bibliography{MajoranaWire}

\end{document}